# Forecasting battery capacity and power degradation with multi-task learning

Weihan Li[1,2,6]*, Haotian Zhang[1], Bruis van Vlijmen[4,5], Philipp Dechent[1,2], Dirk Uwe Sauer[1,2,3]*

## SUMMARY

Lithium-ion batteries degrade due to usage and exposure to environmental conditions, which affects their capability to store energy and supply power. Accurately predicting the capacity and power fade of lithium-ion battery cells is challenging due to intrinsic manufacturing variances and coupled nonlinear ageing mechanisms. In this paper, we propose a data-driven prognostics framework to predict both capacity and power fade simultaneously with multi-task learning. The model is able to predict the degradation trajectory of both capacity and internal resistance together with knee-points and end-of-life points accurately at early-life stage. The validation shows an average percentage error of 2.37% and 1.24% for the prediction of capacity fade and resistance rise, respectively. The model's ability to accurately predict the degradation, facing capacity and resistance estimation errors, further demonstrates the model's robustness and generalizability. Compared with single-task learning models for forecasting capacity and power degradation, the model shows a significant prediction accuracy improvement and computational cost reduction. This work presents the highlights of multi-task learning in the degradation prognostics for lithium-ion batteries.

## INTRODUCTION

Nowadays, lithium-ion batteries (LIBs) are widely used as energy sources in many sectors due to their high energy and power density, low self-discharging rate, low price, and long lifetime. However, similar to many other electrochemical systems, LIBs suffer from both energy and power fade inevitably during usage and storage, which are reflected by the capacity loss and internal resistance increase [1]. Therefore, the ratio of the remaining capacity and the internal resistance compared with their initial values at the beginning of life (BOL) are widely used as state-of-health (SOH) indices in battery management systems (BMSs), i.e., SOH-C and SOH-R [2]. Not only the nowcasting but also the forecasting of the battery degradation is attracting increasing attention considering their applications in more and more fields, e.g., cell design optimization, lifetime prediction, and warranty cost estimation, etc. Apart from the end-of-life (EOL) points, knee points of the degradation trajectory, after which the degradation accelerates, pose challenges for accurate ageing estimation and lifetime prediction. An accurate prediction of the degradation trajectory of each cell with the degradation knee in the battery pack is of crucial importance for both first- and second-life applications, as the lifetime of the whole pack is determined by the worst cell, and the safety issue of one cell can even cause a thermal runaway of the whole battery system. Apart from its significant application in the field, an accurate degradation prediction also saves considerable effort and costs in the characterization of the service life of batteries in the laboratory, which requires a large number of tests at different current levels, cycle depths, temperatures, combinations of idle and operation phases. Running these tests all the way to the end without an accurate degradation prediction not only costs a lot of time, especially for all test cycles that are very similar to the application with many pause times, but also a lot of money. As the degradation of the LIBs is a complex nonlinear process coupled with different physical and chemical mechanisms, e.g., SEI growth [3], lithium plating [4], particle cracking [5], which are affected by multiple stress factors, e.g., temperature, current rate, depth of discharge [6], an early-life stage prediction of the future degradation trajectories of LIBs is challenging. The intrinsic variability caused by minor inconsistencies throughout manufacturing processes will also cause significant degradation differences in the late-life stage under the same real-world operation conditions, which further increases the prediction difficulties [7, 8].

[1] Chair for Electrochemical Energy Conversion and Storage Systems, Institute for Power Electronics and Electrical Drives (ISEA), RWTH Aachen University, Jägerstraße 17-19, 52066 Aachen, Germany
[2] Jülich Aachen Research Alliance, JARA-Energy, Templergraben 55, 52056 Aachen, Germany
[3] Helmholtz Institute Münster (HI MS), IEK-12, Forschungszentrum Jülich, 52425 Jülich, Germany
[4] SLAC National Accelerator Laboratory, Menlo Park, California 94025, United States of America
[5] Stanford University, Stanford, California 94305, United States of America
[6] Lead Contact
*Correspondence: weihan.li@isea.rwth-aachen.de (W.L.), dirkuwe.sauer@isea.rwth-aachen.de (D.U.S.)

Degradation prediction methods can be divided into two main categories: model-based methods and data-driven methods. Model-based methods aim to describe the battery degradation dynamics with a mathematical model, e.g., electrochemical models [9], equivalent circuit models [10] [11], and empirical models [12]. These models describe the physics occurring in the battery at different levels and are developed based on different model parametrization methods. However, it is challenging to model the intrinsic cell degradation variances and to reach a trade-off between model complexity and prediction accuracy. In contrast, data-driven methods aim to extract hidden correlations from battery ageing data in the laboratory or the field and predict the degradation as an end-to-end solution with machine learning. With high-quality battery ageing data, machine learning models are able to extract the features of the ageing patterns, catch the slight variances between cell ageing behaviours in the early-life stage and predict the future degradation trend with high accuracy.

Observing the existing work highlights two main data-driven model categories: the point-prediction models and the curve-prediction models. Point-prediction methods use traditional machine learning and simple neural networks, e.g., linear regression [13, 14], support vector machines (SVMs) [15], Naive Bayes [16], artificial neural networks [17], to predict the EOL point or remaining useful life directly. However, the whole degradation trajectory, together with other important degradation features, e.g., knee-point, is not captured in the case of point-prediction methods. On the contrary, a more sophisticated curve-prediction model aims to predict the future trajectory of the ongoing degradation based on the past data. Most of the curve-prediction models are based on the iterative structure, which uses a sliding window to capture input data from the degradation history and predict the next time-step point. The iterative structure can be realized with different machine learning approaches, e.g., SVMs [18] [19], relevance vector machines (RVMs) [20] [21] or Gaussian process regressions [22] [23]. Although these machine learning models achieved good prediction accuracy, the feature extraction is challenging and relies on expert knowledge. Several features used in the models are also not accessible in a battery management system and therefore limit the application of such models in the real world. In contrast, deep learning models with multiple network layers, e.g., recurrent neural networks (RNNs) [24] and convolutional neural networks (CNNs) [25], learn the features directly from the data. However, the major disadvantage of the iterative approaches is the high computing cost, as the models need to be run iteratively to get the prediction of the whole degradation curve. To solve this problem, Li et al. [26] proposed a one-shot battery degradation trajectory method with deep learning and reduced the computing cost by 15 times compared with the iterative approach. However, there is very limited work in the literature regarding the power degradation prediction [27–31] as previously the interest in mainly consumer electronics and passenger electric vehicles diverted attention to capacity degradation. In such a use case, like passenger electric vehicles, because of the relatively low discharge current rate requirement, capacity is directly related to the driving range. With the start of the wide electrification of the transportation sector, power fade may cause critical performance degradation in some applications, e.g., plug-in hybrid vehicles, heavy-duty vehicles, trains and aircraft. Even for electric passenger vehicles, increasing attention is paid to the negative implications of the power degradation with respect to internal resistance increase on fast charging [32] and thermal management [33]. The battery cells with identical capacity fade may have entirely different capacity EOL due to the difference in power fade, vice versa. Therefore, an accurate prediction of both capacity and power fade in the early-life stage is necessary for a safe and reliable battery system in both first- and second-life applications. RVMs and CNNs have been used in the joint prediction of the resistance increase and capacity fade successfully [30, 31]. However, these deep learning models were based on the measured voltage, current and SOC data under the strict requirement of a standardized, constant current charging and discharge profile, limiting the application of these methods in the field. Furthermore, the implications of the feature selection on the degradation prediction performance is unclear.

In this work, we aim to bridge the aforementioned research gap and explore a prognostics framework for capacity and power degradation trajectory prediction with the multi-task learning (MTL) and sequence-to-sequence (S2S) models. We first present the concept of knee points and EOL points in first- and second life for not only capacity degradation but also power degradation. We then present the MTL framework based on the S2S model for capacity and resistance trajectory prediction with regularization to improve the model's generalizability. Based on the training on the entire degradation data, the MTL model is able to project the capacity and power degradation trajectories simultaneously with as few as 100 cycles of data, together with the knee points and EOL points. The model was found to be accurate and robust even under capacity and resistance estimation noise. The multi-task nature of the model leads to better prediction accuracy and a reduction of 50% in computing times compared with single-task learning (STL) models in forecasting capacity and power degradation. Overall, the high accuracy and robustness of the MTL model highlight the effectiveness of multi-task learning in degradation prediction. The method in this work can be applied broadly for other cell chemistry and other energy storage systems.

## AGEING DATA ANALYSIS

The battery degradation dataset used in this work was obtained from the cyclic ageing test of 48 Sanyo/Panasonic 18650 NMC/Graphite LIBs with the goal to discover the influence of the intrinsic manufacturing variations on the cells' lifetime [7, 26]. The cells have a nominal capacity of 1.85 Ah and were tested under the same ageing scenario at 25°C. In the cycling tests, the cells were charged/discharged under the constant current constant voltage (CC-CV) regime with 2 C for 30 min between 3.5 V and 3.9 V. There are on average 17 characterization tests for each cell and about 160 charging-discharging cycles between each two characterization tests, where the capacity checkup under different current rates and hybrid pulse-power characterization tests were conducted. In this work, the capacity at a discharge current of 1 A and the 2-second resistance at a 2 A charging pulse around 90% SOC are used as the index for SOH-C and SOH-R, respectively. The power fade of the battery cells is usually determined by the cell impedance. The cells were cycled beyond the standard industry-level EOL to get insights into the full-lifetime performance of the LIBs regarding both capacity and power degradation. After obtaining the raw data, the measurement of remaining capacity and internal resistance were interpolated to create the supervised learning dataset, using the piecewise cubic hermit interpolating polynomial (PCHIP) function in MATLAB. As shown in Figure 1(a), the EOL regarding capacity degradation in the first- and second-life are defined as 80% (EOL80) and 65% (EOL65) of the nominal capacity. Similarly, the EOL regarding power degradation in the first- and second-life are defined as 120% (EOL120) and 130% (EOL130) of the average initial resistance, as shown in Figure 1(b). The values of the capacity and resistance at the EOLs for the first- and second-life are chosen based on the knowledge of the ageing data and can be adapted in different applications or use cases, e.g., EOL for power fade can also be defined as 200% of the initial resistance in some cases. The knee-points of both capacity and power degradation are identified based on the curvature of the degradation trajectories and indicate the accelerated degradation speed. As the cells were tested under the same ageing scenario to explore the intrinsic cell manufacturing variability, the variance in the degradation trends is very small at the BOL and starts to increase significantly in the mid-life, which increases the challenge for an accurate early-life cell-specific degradation trajectory prediction. Figure 1(c) shows the distribution and the pairwise relationships of three capacity degradation metrics. The battery cycle life at the capacity knee-point shows a strong correlation with EOL80, which makes sense that earlier knee causes earlier onset of rapid degradation, i.e., short cycle life. In contrast, the remaining capacity at the capacity knee-point shows a poor correlation with EOL80, highlighting that the remaining capacity at which rapid degradation starts is not indicative of cycle life. This may further indicate that the accelerated degradation is not a consequence of a decrease in capacity, rather internal phenomena in the battery. Compared with capacity degradation, the power degradation of the cells shows similar features, as shown in Figure 1(c). Although the correlation between the battery life cycle at the knee-point and EOL120 is strong, similar to that in capacity degradation, the poor correlation between the knee-point resistance and EOL120 further demonstrates the nonlinear relationship between the degradation state and degradation knee. The pairwise correlation and distribution of all the metrics in capacity and power degradation can be further found in Figure S1 and Figure S2. The strong correlation between four EOLs might suggest that the degradation of these cells is limited by the SEI growth, as the resistance is the dominating factor determining capacity lifetime. Figure S3 and Figure S4 highlight the highly unclear correlation of the capacity of the cells at the BOLs with their final EOL cycle numbers, whereas the same correlation for the internal resistance is higher. This indicates that the initial states of cell resistance hold slightly more predictive information about the resistance states at EOL compared to the capacity measurements, which might yield better prediction accuracy as well. In summary, similar features in capacity and power fade highlight the nonlinear relationship between the capacity fade and resistance rise under a specific operation mode. Although this correlation is nonlinear and may change with battery materials and operation modes influenced by both cyclic ageing and calendric ageing [34], the information of the capacity degradation may theoretically benefit the prediction of the power degradation trajectory, vice versa. However, the degradation trajectory of capacity and power are in different directions with different shapes, the co-prediction of which is challenging.

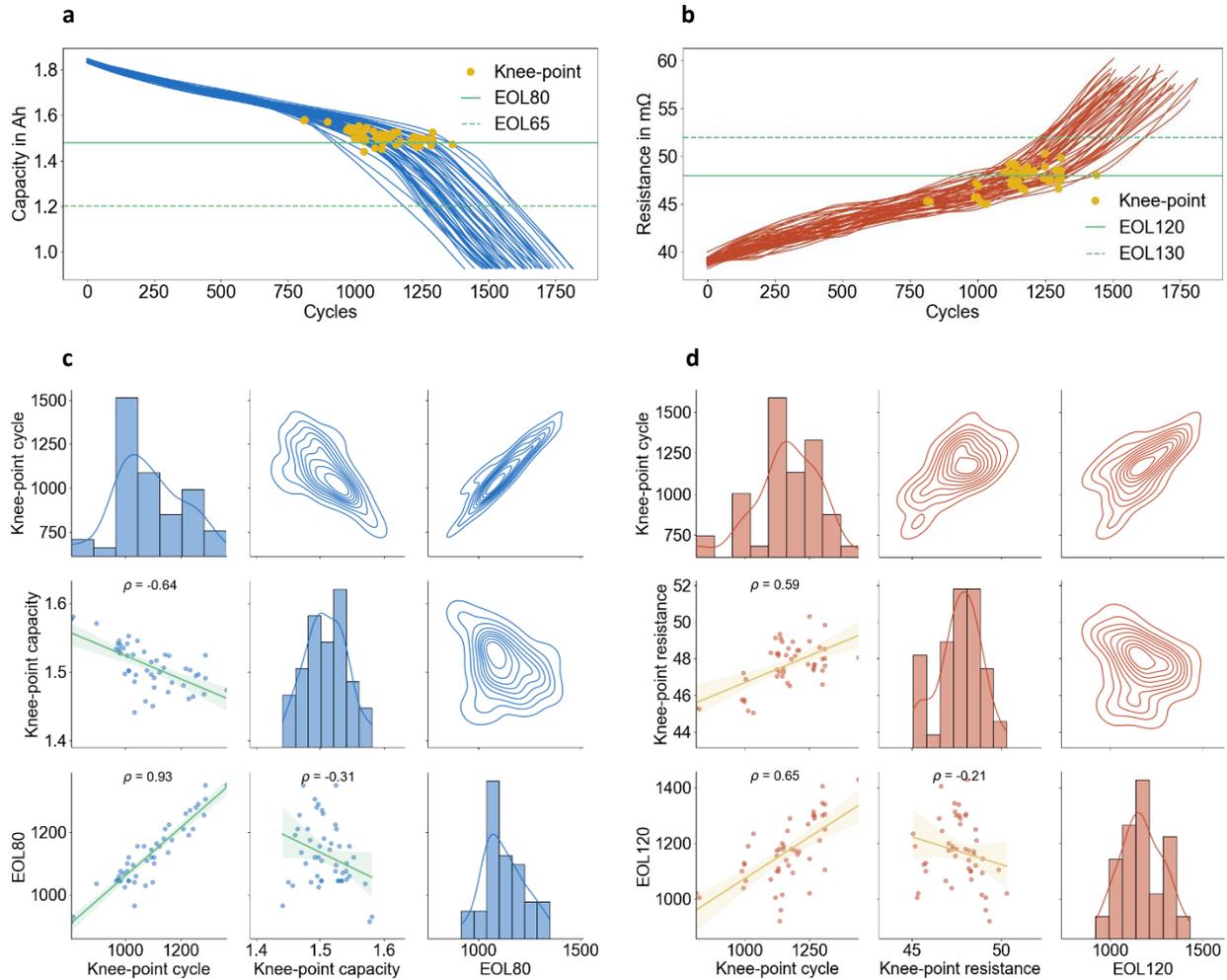

**Figure 1.** Analysis of the ageing dataset: (a) Capacity degradation curves and (b) internal resistance increase curves of 48 cells. (c) Distribution and pairwise relationships of three capacity degradation metrics: Battery cycle life at capacity knee-points; remaining capacity at capacity knee-points; battery cycle life at EOL80. The Pearson correlation coefficient is represented by $\rho$. (d) Distribution and pairwise relationships of three power degradation metrics: Battery cycle life at resistance knee-points; relative internal resistance at resistance knee-points; battery cycle life at EOL120.

**PROGNOSTICS FRAMEWORK**

The framework of the data-driven degradation prognostics for LIBs is shown in Figure 2(a), which can be split into two main parts, namely the initial training of the model and the deployment and updating of the model in operation. The training data consists of both the capacity and internal resistance time-series data of a specific cell type either measured with experimental ageing tests in the laboratory or calculated with the SOH estimation models in BMSs under operation. The training data is stored in a cloud server for model training, where the bidirectional connection between the cloud server and all inference devices in the laboratory and field enables the update of the prognostics model parameters over the air. With the prognostics framework, the digital twin of the battery cells [10] in the cloud is able to forecast both the capacity and power degradation trajectory of each cell or pack together with the EOL for first- and second-life applications and degradation knee-points. The information of the capacity and power fade status of each cell can be logged with the digital cell-health passport along with other operating environment information and stress factors, which can be used as inputs in the MTL model for multi-feature co-prediction of the degradation. Overall, the whole framework enables the tracking of both capacity and internal resistance development and the monitoring of capacity and power degradation of the battery cells, which increases the safety and longevity of the battery systems in both first- and second-life.

MTL is a machine learning approach to solve multiple related learning tasks at the same time, which improves the model generalization by adding relative tasks in the learning process and exploiting commonalities and differences across tasks using a shared representation [35]. The goal of this work is to build an MTL model to predict the capacity and power degradation based on the past degradation data at the same time to improve both computation efficiency and prediction accuracy for the task-specific models. Figure 2(b) highlights the general architecture of the MTL model and STL models. The MTL model used in this work is based on the hard parameter sharing structure, where the shared parameters can be directly affected by both the capacity and internal resistance prediction tasks during training. The training of the MTL model leads to the determination of a representation to capture both of the prediction tasks, which helps to avoid over-fitting. In contrast, the STL models have separate model parameters which will not be influenced by each other.

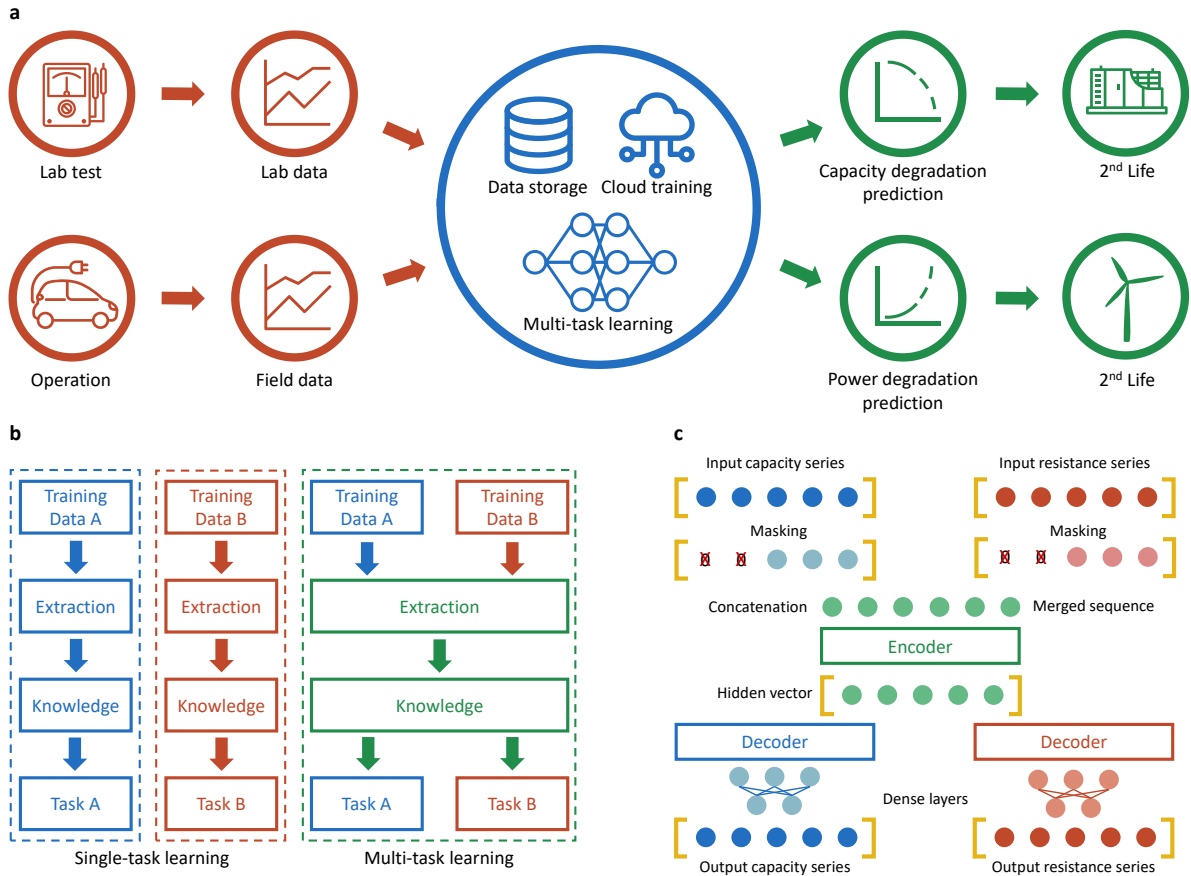

**Figure 2.** Battery capacity and power degradation prognostics: (a) an overview of the prognostics framework. (b) Representation of the architecture of multi-task learning and single-task learning. (c) The architecture of the multi-task sequence-to-sequence learning model.

The input data to the MTL model is the arrays of remaining capacity and internal resistance, from the BOL of the cell to the present time point. The outputs of the MTL model are the future trajectories of both remaining capacity and internal resistance. As the lifetime of the cells varies with each other due to intrinsic manufacturing variances and operation condition variances, the length of the outputs is uncertain. With more historical data available, the sizes of both the input and output series will also change with time. To predict the capacity and power degradation trajectories at any time point without iterative calculation and re-training, we use an RNN-based S2S learning model [36] in this work. Figure 2(c) shows the architecture of the MTL model based on S2S learning, where the input approaches are increasing windows of the past capacity array and resistance array, from the BOL of the cell to the current time point. Similarly, the outputs are decreasing windows of the future capacity array and resistance array starting from the current time point. The encoder generates a hidden vector representing the input sequence, and the decoders regenerate the output sequences based on the hidden vector (See 'Methods' for details of the model structure and model training). To validate the performance and robustness of the degradation prediction model when performing in real-life operation conditions, two scenarios were chosen, i.e., validation with ideal inputs, as well as noisy inputs with degradation diagnosis errors. As introduced in 'Methods', 20% of the whole dataset (10 cells) which are not used in the training phase were used as the test data.

## RESULTS AND DISCUSSION

### Degradation Prediction from Normal Input

Figure 3 shows the results of capacity degradation prediction from normal input. From the results, the model is able to predict the capacity degradation trajectory accurately. The predictions are accurate even with 100 cycles of input data, as the initial values of the capacity mean absolute percentage error (MAPE) in the best-case and worst-case is 1.1% and 3.0%, respectively. Figure 3(a) shows the predictive ability of the model in the best-case in early-life, mid-life and late-life. Similarly, Figure 3(b) shows the curves from the worst-case cell. The performance of the capacity degradation curve prediction remains to have a good performance in different phases of the cells' life, which are below 6%. This is also confirmed by Figure 3(c), which shows the progression of the MAPE of the predicted capacity degradation curves over the cell's lifetime for the best- and worst-cases. The median values of the mean and maximum capacity MAPE of all the test cells are 2.3% and 5.4%, respectively (see Figure S5 for more details). Figures 3(d), 3(e), and 3(f) show the progression of the three highlighted capacity degradation metrics of interest, namely the EOL80 error, EOL65 error, and capacity knee-point error. In most cases, the best-case shows a lower error compared with the worst-case. As is expected with forecasting models, our model's predictions of these metrics become increasingly more accurate as the model receives more input data. This trend is more evident in the worst-case cell than in the best-case, where the prediction errors fluctuate in a very low-error area.

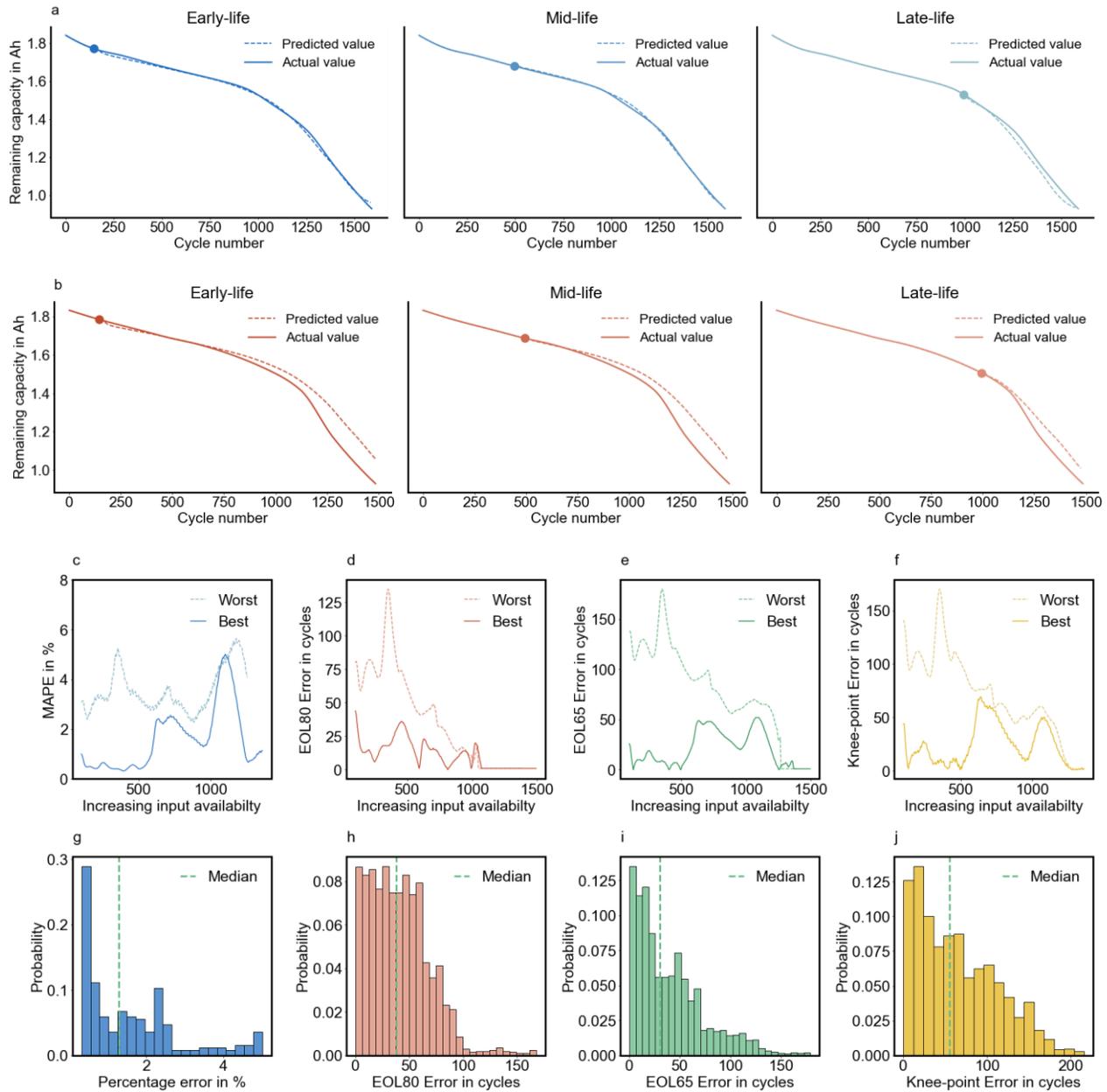

**Figure 3.** The capacity fade prediction performance of the MTL model under the normal condition: Figures (a) and (b) show the predicted future capacity fade curves for the worst- and best-case in early-, mid-, and late-life conditions. Figure (c) shows the progression of the MAPE with the increase of the input sequence length. Figures (d), (e), and (f) describe the best- and worst-case progression for the EOL80, EOL65, and the knee-point absolute errors, respectively. Figure (d) shows the distribution of the curve MAPE for the best case. Figures (h), (i), and (j) show the distribution of the EOL80, EOL65, and knee-point absolute error metrics for all test cells.

Figure 3(g) shows the histogram of the errors in prediction for all the curves over the lifetime of the best-case cell, with the median error of 1.3%. Figures 3(h), 3(i), and 3(j) depict the distribution of the EOL80 error, EOL65 error, and knee-point error for all the test cells, respectively, with the median values of which are 38, 33, and 55 cycles. The model shows good performance at all points of interest for capacity degradation prediction, as is clear from the metrics analysis.

As shown in Figure 4, the results demonstrate that the model also has a good performance for power degradation prediction. Figures 4(a) and 4(b) depict the predicted resistance increase curve in early-, mid-, and late-life for the best- and worst-case, respectively. For the worst-case, Figures 4(c) and 4(d) show the progression and histogram of the MAPE of the predicted resistance increase curves over the cell's lifetime, with the initial and median resistance MAPE as 1.8% and 1.5%, respectively. For the best case, Figures 4(c) and 4(g) show a similar progression and histogram of the resistance increase prediction MAPE over the lifetime with the initial and median values as 1.1% and 0.8% (see Figure S5 for more details). Compared with the capacity fade prediction, the power fade prediction shows smaller errors, especially in the early-life, which is maybe due to the higher correction of the internal resistance at early-life with the EOL, as discussed in Section 2. The median values of the mean and maximum resistance MAPE of all the test cells are 1.2% and 2.5%, respectively. Figures 4(d), 4(e), and 4(f) further show the progression of the EOL120 error, EOL130 error, and resistance knee-point error. For most cases, the best-case shows a lower error than the worst-case. However, in the EOL120 metric, the best-case shows a similar accuracy compared with the worst-case, which is also possible as the best- and worst-case are referring to the average curve prediction errors. Figures 4(h), 4(i), and 4(j) depict the distribution of EOL120 error, EOL130 error, and resistance knee-point error for all the test cells, respectively, with the median values as 41, 34, and 61 cycles. Based on the validation results, the MTL model can accurately predict both the capacity and power degradation trajectory, even with only 100 cycles of data. In most conditions, the MTL model shows an accuracy increase with more input data available. However, the prediction accuracy of some metrics decreased at some time points, which can be attributed to the model having few meaningful points to predict and the errors between those few points being amplified. This is further demonstrated by Figures S9–S16, where the typical shape of each degradation metric prediction error between certain cells is similar to each other, i.e., the prediction errors increase or decrease at the same time points when new data is available. The dependence of prediction error on the size of input points likely varies with the certain distinct regions of the degradation curve, such as the initial fade, linear degradation section and the section around the knee point. The exact dependence warrants further investigation.

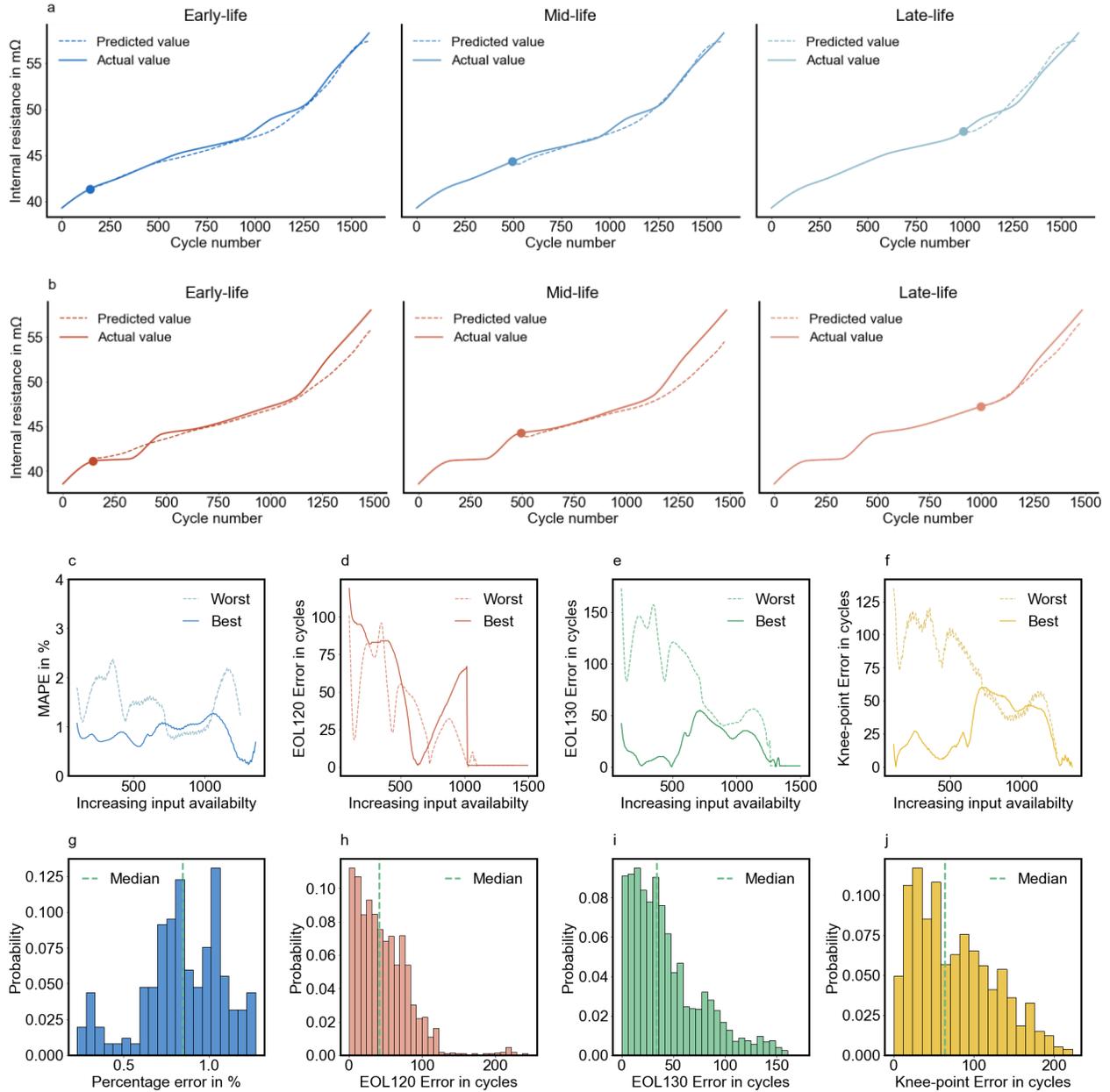

**Figure 4.** The power fade prediction performance of the MTL model under the normal condition: Figures (a) and (b) show the predicted future internal resistance increase curves for the best- and worst-case in early-, mid-, and late-life conditions. Figure (c) shows the progression of the MAPE with the increase of the input sequence length. Figures (d), (e) and (f) describe the best- and worst-case progression for the EOL120, EOL130, and the knee-point absolute errors, respectively. Figure (g) shows the distribution of the curve MAPE for the best case. Figures (h), (i), and (j) show the distribution of the EOL120, EOL130, and knee-point absolute error metrics for all test cells.

**Degradation Prediction under Diagnosis Errors**

The noisy input condition was used to demonstrate the robustness of the MTL model under the capacity and resistance estimation errors, which are not avoidable in BMSs. The noisy condition was realized by adding a zero-mean white Gaussian noise, which was configured with the standard deviation (σ) of 1% of its initial point value to both the capacity and resistance input sequences. With this white Gaussian noise, the maximum deviation is about 3% of the actual values, which matches with the industrial requirement of the maximum estimation error for SOH estimation. The prediction performance of the MTL model for capacity and power degradation from noisy input sequences are shown in Figure 5 and Figure 6, respectively.

For both capacity and power degradation prediction, the addition of noise to the input only has a small effect on the mean accuracy of the model for both capacity and power degradation prediction. For the capacity degradation prediction, Figures 5(a) and 5(b) show the predicted curves for the worst- and best-case in early-life, mid-life, and late-life. The initial and median capacity MAPEs in the worst-case are 3.3% and 3.7%, respectively, as shown in Figure 5(c), whereas in the best-case, the initial and median capacity MAPEs are 0.9% and 1.3%, respectively, as shown in Figures 5(c) and 5(g). Figures 5(d), 5(e), and 5(f) depict the progression of the EOL80 error, EOL65 error, and capacity knee-point error for the best- and worst-case. The overall trend remains the same as the normal condition, but the error curves are not smooth. Figures 5(h), 5(i), and 5(j) show the distribution of EOL80 error, EOL65 error, and capacity knee-point error for all the test cells with the median values as 39, 34, and 61 cycles. The median value of the mean and maximum capacity curve MAPE for all test cells is 2.5% and 6.6%, respectively. For the power degradation prediction, the predicted curves for the best- and worst-case at three life-stages are shown in Figures 6(a) and 6(b). Figure 6(c) shows the progression of the resistance curve prediction MAPE with the initial MAPE for the best- and worst-case are 1.1% and 1.0%, respectively. Figures 6(d), 6(e), and 6(f) depict the progression of the EOL120 error, EOL130 error, and resistance knee-point errors for the best- and worst-case, which remains the same decreasing trend with more data available. Figures 6(g) shows the histogram of the resistance curve MAPE with the median values for the best-case as 0.9%. Figures 6(h), 6(i), and 6(j) show the distribution of the EOL120 error, EOL130 error, and resistance knee-point error metrics with the median values as 39, 35, and 70 cycles. The median value of the mean and maximum resistance curve MAPE for all the test cells is 1.2% and 3.3%, which are lower than those of capacity degradation prediction.

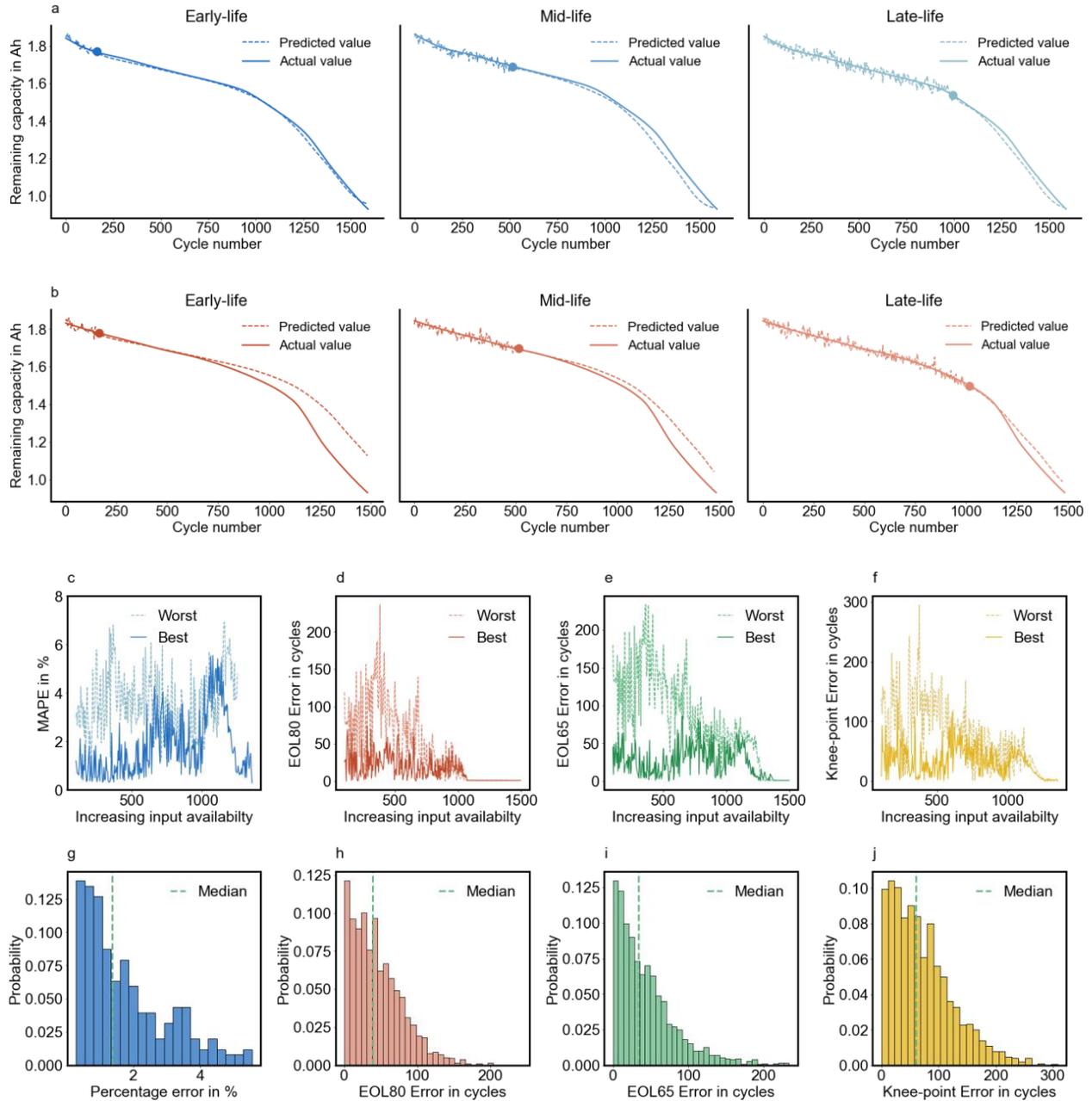

**Figure 5**. The capacity fade prediction performance of the MTL model under a noisy input condition: Figures (a) and (b) show the predicted future capacity fade curves for the best- and worst-case in early-, mid-, and late-life conditions. Figure (c) shows the progression of the MAPE during the input sequence length increase. Figures (d), (e) and (f) describe the best- and worst-case progression for the EOL80, EOL65, and the knee-point absolute errors, respectively. Figure (g) shows the distribution of the curve MAPE for the best case. Figures (h), (i), and (j) show the distribution of the EOL80, EOL65, and knee-point absolute error metrics for all test cells.

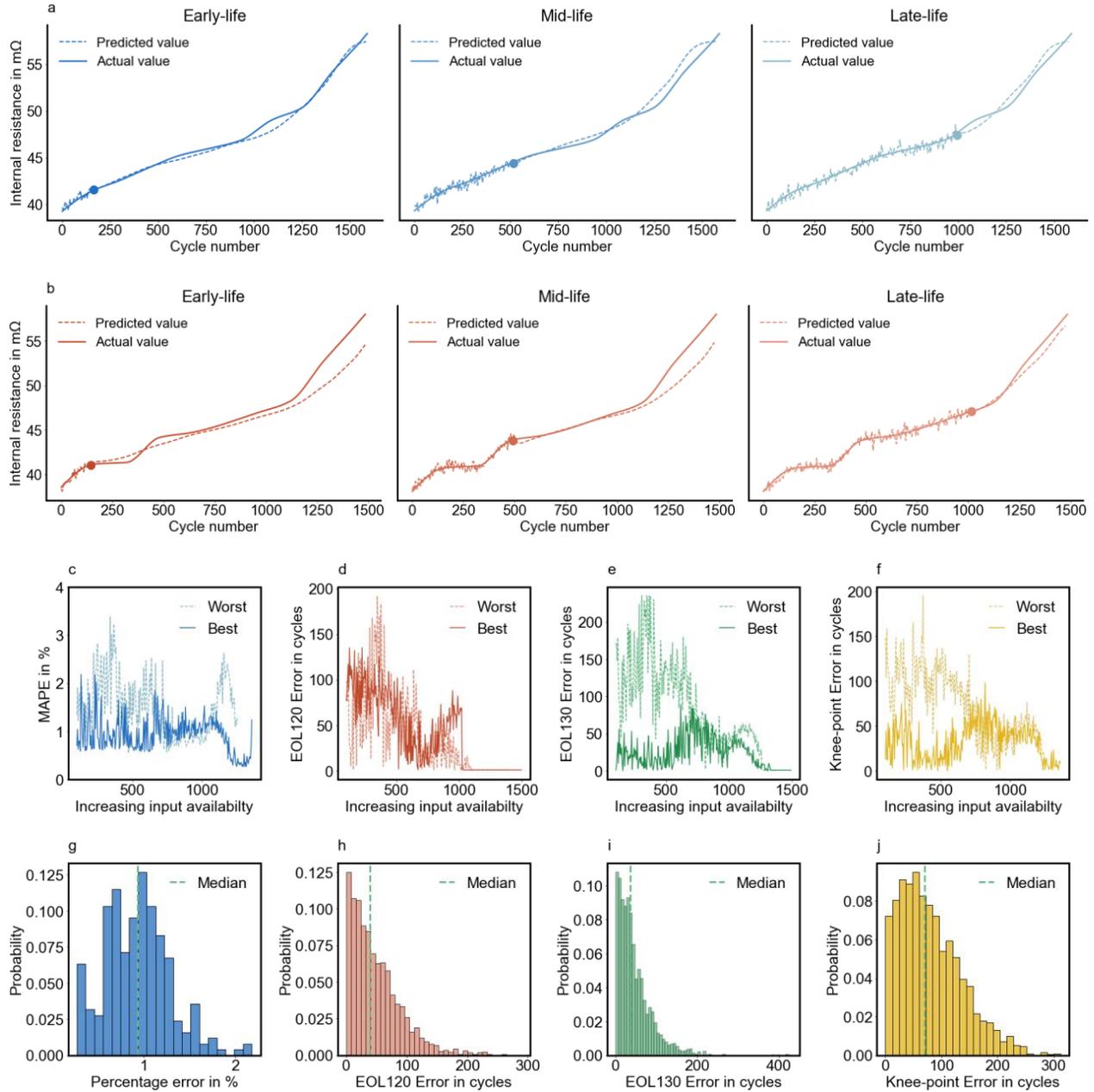

**Figure 6**. The power fade prediction performance of the MTL model under a noisy input condition: Figures (a) and (b) show the predicted future internal resistance increase curves for the best- and worst-case in early-, mid-, and late-life conditions. Figure (c) shows the progression of the MAPE during the input sequence length increase. Figures (d), (e), and (f) describe the best- and worst-case progression for the EOL120, EOL130, and the knee-point absolute errors, respectively. Figure (g) shows the distribution of the curve MAPE for the best case. Figures (h), (i), and (j) show the distribution of the EOL120, EOL130, and knee-point absolute error metrics for all test cells.

Based on the results, the MTL model is stable to handle capacity and resistance estimation errors in BMSs without significant overall performance degradation (see Figure S8 for more details). Even though the input sequence contains noisy signals, the model's output is smooth. In most cases, the best and worst-case cells remain the same after the noise signal is added. These phenomena show that the model's encoder part can extract the features and partly signals. However, the noisy input clearly affects the prediction accuracy, especially on the maximum MAPE metric, which is expected since the input is not ideal. However, the maximum MAPE only appears in one cell at some time points as outliers. The overall trend remains the same as the normal condition, and the median value of the metrics only shows a slight increase. These phenomena indicate that the MTL model can predict both capacity and power degradation with high accuracy over the whole life. The primary metrics of the MTL model are summarized in Table. 1.

To further analyze the model robustness under different scales of diagnosis errors, four more different noises with standard deviations were added to the normal data in both capacity and resistance arrays. As shown in Figures 7(a) and 7(c), the mean and median errors of the trajectory predictions for capacity fade and power fade are very robust and show little change with the increase of the input noise. Only the maximum prediction error is influenced by the noise and increases with the standard deviation of the noise. The median prediction errors of the EOLs and knee points for capacity fade and power fade are also stable under different noises, as shown in Figures 7(b) and 7(d). To further enhance the model performance and reduce the maximum prediction errors, various filters can be implemented to smooth the input data. This work also highlights the significance of accurate battery diagnosis algorithms for degradation prediction and can be used to select battery diagnosis algorithms for different use cases considering the compromise between the model cost and prediction performance. The MAPEs and mean absolute errors (MAEs) for the curve prediction and the absolute errors for the point predictions are summarized in Table 1.

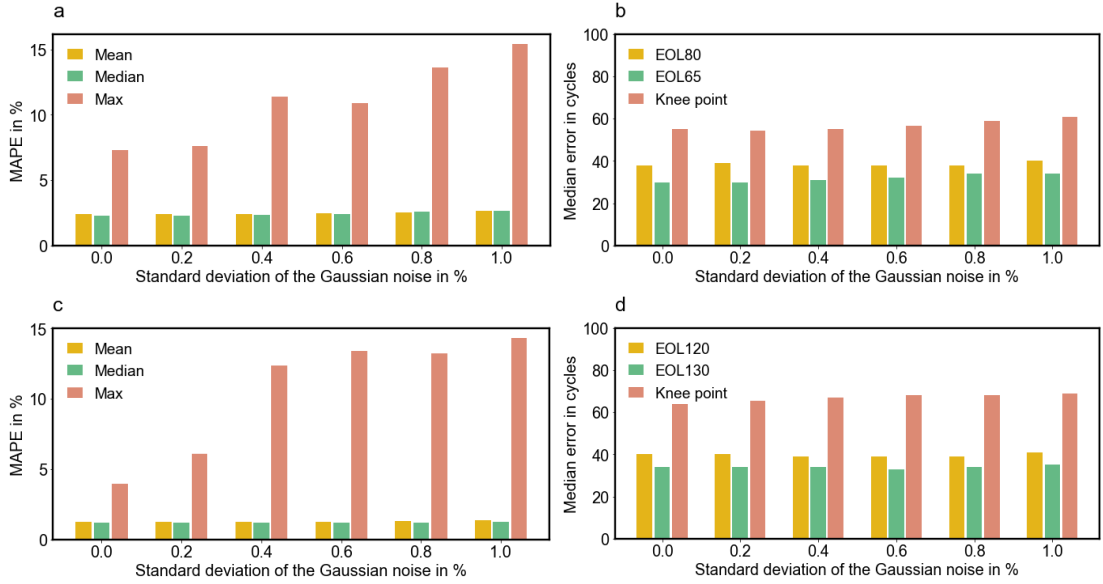

**Figure 7**. The robustness of the MTL model under different diagnostics errors: Figures (a) and (b) show the model performance in the prediction of the capacity fade trajectory and key degradation metrics, i.e., EOL80, EOL65 and knee point. Figures (c) and (d) show the model performance in the prediction of the power fade trajectory and key degradation metrics, i.e., EOL120, EOL130 and knee point.

|  | Metric | Normal | 0.2% Noise | 0.4% Noise | 0.6% Noise | 0.8% Noise | 1% Noise |
|---|---|---|---|---|---|---|---|
| Capacity | Mean curve MAPE [%] | 2.37 | 2.39 | 2.41 | 2.46 | 2.54 | 2.65 |
|  | Max curve MAPE [%] | 7.30 | 7.59 | 11.37 | 10.92 | 13.65 | 15.40 |
|  | Median curve MAPE [%] | 2.27 | 2.30 | 2.36 | 2.41 | 2.58 | 2.52 |
|  | Mean curve MAE [mAh] | 28.4 | 28.6 | 28.9 | 29.5 | 30.5 | 31.82 |
|  | Max curve MAE [mAh] | 87.2 | 90.7 | 145.2 | 139.8 | 177.4 | 199.36 |
|  | Median curve MAE [mAh] | 27.4 | 27.9 | 28.6 | 29.2 | 31.9 | 31.8 |
|  | Median knee-point error [cycle] | 55 | 55 | 55 | 57 | 59 | 61 |
|  | Median EOL80 error [cycle] | 38 | 39 | 38 | 38 | 38 | 40 |
|  | Median EOL65 error [cycle] | 30 | 30 | 31 | 32 | 34 | 34 |
| Resistance | Mean curve MAPE [%] | 1.24 | 1.25 | 1.27 | 1.28 | 1.30 | 1.34 |
|  | Max curve MAPE [%] | 3.99 | 6.07 | 12.39 | 13.42 | 13.25 | 14.31 |
|  | Median curve MAPE [%] | 1.22 | 1.22 | 1.22 | 1.21 | 1.22 | 1.23 |
|  | Mean curve MAE [mΩ] | 0.64 | 0.64 | 0.65 | 0.66 | 0.67 | 0.69 |
|  | Max curve MAE [mΩ] | 2.32 | 3.03 | 6.09 | 6.55 | 6.48 | 0.69 |
|  | Median curve MAE [mΩ] | 0.63 | 0.63 | 0.63 | 0.62 | 0.61 | 0.62 |
|  | Median knee-point error [cycle] | 64 | 66 | 67 | 68 | 68 | 69 |
|  | Median EOL120 error [cycle] | 41 | 40 | 39 | 39 | 39 | 41 |
|  | Median EOL130 error [cycle] | 34 | 34 | 34 | 33 | 34 | 35 |

**Table 1.** Validation results for capacity and power degradation prediction from normal input and noisy input with different levels of noise for the MTL model.

**Comparison with Single-Task Learning Models**

As a benchmark, STL models for capacity and power degradation prediction were designed and trained (see Methods for the details of the STL models and Table S2 for the training parameters of the STL models) on the same dataset to be compared with the proposed MTL model. The benchmark STL models consist of four bidirectional LSTM layers with 64 hidden nodes in both encoder and decoder for fairness. The two STL models were trained with the capacity and resistance prediction MAE as the loss function with a learning rate of $1 \times 10^{-4}$ until the convergence.

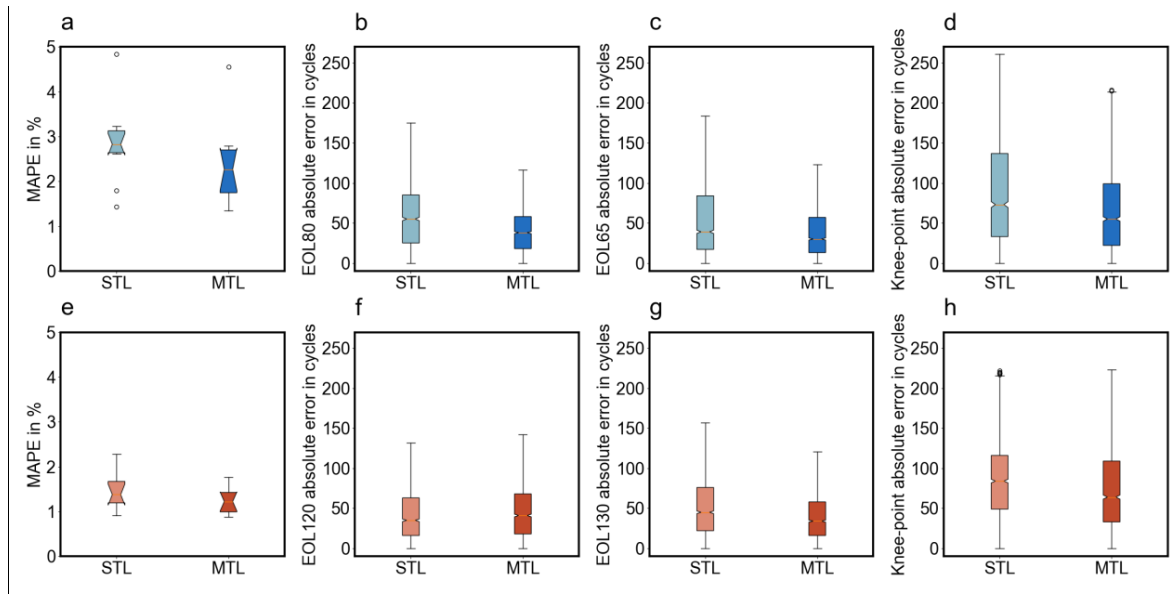

**Figure 8**. Comparison results between the STL models and the MTL model. Figure (a) shows the distributions of capacity curve MAPE. Comparison plots for the STL model and the MTL model for the (b) EOL80 error, (c) EOL65 error, and (d) capacity degradation knee-point error are shown in the subsequent three figures. Figure (e) shows the distributions of resistance curve MAPE. The following three figures show the comparison between the STL model and the MTL model for the (f) EOL120 error, (g) EOL130 error, and (h) power degradation knee-point error.

The comparison results are shown in Figure 8 for both capacity and power degradation prediction. Figure 8(a) shows the performance of the STL model and the MTL model in capacity curve prediction, where the mean, median and maximum MAPE of the MTL model are all lower than those of the STL model, highlighting the prediction accuracy improvement with the MTL. Similar results can be seen for the points of interest in capacity degradation, as shown in Figures 8(b), 8(c), and 8(d), where the confidence interval (CI) ranges of the MTL model are also narrower than those of the STL model, demonstrating that the prediction uncertainty decreases with the MTL. Similarly, the prediction performance of the STL model and the MTL model for the resistance curve and three related metrics, i.e., EOL120, EOL130, and resistance knee-point, are shown in Figures 8(e), 7(f), 7(g) and 7(h). Apart from the EOL120, where the prediction accuracy of the MTL model is slightly lower than that of the STL model, all the other metrics show a better performance with the MTL model. In general, the prediction accuracy of all the metrics for capacity and power degradation was increased due to the structure of the MTL model, where both the capacity sequence and resistance sequence data are used as input for a better prediction of the future degradation trends, no matter of capacity or power. The improvement of the generalization by using the domain information contained in the training data of capacity fade and power fade as inductive bias enhances the model performance in both tasks. With more features available based on the MTL architecture, the prediction accuracy of almost all the metrics in both capacity and power degradation prediction was increased. Interestingly, not only the MTL model but also the STL models show a lower prediction error for power fade compared with capacity fade. This further verifies the influence of the correlation between early-life and late-life degradation on lifetime forecasting. Furthermore, the distributions of the computational cost of the STL models and the MTL model for each prediction are shown in Figure 9. As expected, the MTL model reduces about 50% of the computational cost compared with the STL models, which is contributed by the fewer model parameters of the MTL model compared with the sum of the parameters in the STL models. Furthermore, the reduction of the preprocessing time also increases the computational efficiency of the MTL model. The shared representations used by the MTL model highlight the increase of both data efficiency in training and prediction efficiency in implementation based on the MTL architecture.

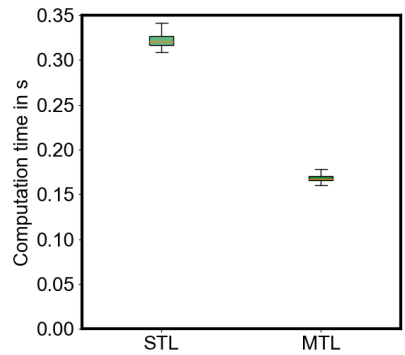

**Figure 9**. Comparison of the computational cost between the STL models and the MTL model.

**Model Performance under Other Datasets**

The proposed MTL model can be used to predict the trajectories in capacity fade and power fade due to both intrinsic manufacturing variances and external operation stress factors. In order to further validate the model performance in degradation prediction under varying external stress factors, an MTL model and two STL models were trained and validated with the dataset from [13], where 124 LFP/graphite battery cells were aged under different charging scenarios. The details of the dataset and the validation results, as shown in Figures S17 to S20, further highlights the high performance of the MTL model in both capacity fade and power fade prediction under very large degradation variances. Despite the varying charging conditions throughout this dataset, the MTL model was still able to achieve good prediction accuracy, namely mean capacity curve MAPE of 1.54% and mean resistance curve MAPE of 1.62%.

**Applications and Outlook**

Future work in the space of this work is broad. Although we provide the proof-of-concept, the performance of the MTL in dealing with multi-feature co-prediction problems can be further validated with not only the experimental datasets with more significant variances in stress factors but also the data from the field with mixed operation conditions. The structure of the proposed MTL model makes it possible to migrate easily between different datasets or battery cells with fine-tuning, further increasing the data efficiency and model generalization ability.

We hope this work also inspires the development of battery lifetime prediction models for multi-prediction of other health indicators, i.e., EIS spectrum [37], ultrasonic signal [38], and pressure signal [39], which can be measured by modern BMSs with new sensors. As the shapes of the trajectory of these new indicators are still unclear and may vary with each other on a large scale, the co-prediction of them will be challenging. With the ability to forecast the multi-features of the battery cells, the MTL model can help to increase the accuracy of lifetime prediction, which further increases the reliability of the battery system in the field. The knowledge of the EOLs and knee-points in both capacity fade and power fade enables a better plan of predictive maintenance, especially for the second-life applications, reducing the total cost of the ownership and introducing exciting developments in battery warranties and insurance. With the prediction model presented here, only a small part of the tests need run to the end in the laboratory in the future, while for the large part of the tests, it will be enough to run a few hundred cycles and then predict the expected service life with the help of our model. This not only saves considerable effort and costs but will also accelerate the speed of innovation through faster feedback and give suppliers of products with batteries a much higher degree of certainty with regard to the expected and guaranteed service life. Additionally, we recommend applying this model in capacity and power prediction at the battery pack level and to other cell chemistry or even other electrochemical systems, i.e., fuel cells and super-capacitors. Finally, we hope that this work inspires more and more applications of deep learning models in battery lifetime prediction, and we encourage the use of interpretable machine learning models to discover the complex correlations of the degradation trajectories of different features.

**CONCLUSIONS**

In this work, we designed a prognostics framework for battery capacity and power degradation prediction with multi-task learning. Using the past capacity and resistance sequence as our only source of inputs without further feature engineering, we trained our multi-task sequence-to-sequence learning model based on the ageing dataset of 48 NMC/graphite cylindrical cells. With the multi-stage training strategy and regularization, the model converges in both capacity and power degradation prediction without overfitting. The model provides the prediction of the entire future capacity and power degradation trajectories in one shot from as few as 100 cycles. Not only the endpoints of both first- and second-life but also the degradation knee-points are predicted together with the trajectories for capacity and power degradation. The model shows an average mean absolute error of 2.37% and 1.24% for capacity and power degradation prediction, respectively, making it on par with industry standards. We also validate the model with various levels of noisy input data, and the results demonstrate the robustness of the model under capacity and resistance estimation errors in battery management systems. Compared with the single-task learning models, the model not only shows a higher performance for the prediction of most of the degradation metrics but also reduces the computational cost by roughly 50%. Furthermore, the model framework showed the ability to be re-trained on a larger dataset with varying charging conditions while retaining the ability to make accurate predictions. In summary, the multi-task prognostics framework presented in this work provides an end-to-end solution for accurate multi-feature battery lifetime prediction with deep learning, which is worth following future work to expand the application to more electrochemical systems and new appearing degradation features.

**EXPERIMENTAL PROCEDURES**

**Resource Availability**

*Lead Contact*

Further information and requests for resources and materials should be directed to and will be fulfilled by the Lead Contact, Weihan Li (weihan.li@isea.rwth-aachen.de).

*Material Availability*

This study did not generate any unique material.

**Data and Code Availability**

The raw experiment dataset of 48 cells used in this study are available online at https://doi.org/10.18154/RWTH-2021-04545. The pre-processed experiment dataset of 48 cells, the code for experiment data analysis and the code for the deep learning models are available online at https://git.rwth-aachen.de/isea/battery-degradation-trajectory-prediction.

**Knee-Point Identification**

Knee point is a critical metric in the battery degradation process, which can be commonly observed [40–43]. The degradation knee points in this work are defined as the points of maximum curvature of the degradation trajectories and can be identified offline based on the method from Satopaa et al. [44]. Before the identification of the knee points in either capacity degradation or power degradation curves, the data need to be smoothed. The normalisation of the degradation data into a range between 0 and 1 is first carried out and the normalized curve is then transferred to a concave and increasing form to simplify calculation process. The maximum difference between the normalized x and y values can be determined and the position of this point is denormalized to obtain the cycle number of the knee point. Figures S5(a) and S5(b) visualize this offline knee-point identification process based on the capacity degradation data. With this method, the value of the degradation gradient at the knee-point can be determined and the corresponding point on the gradient curve of the predicted degradation curve can be obtained. This approach enables the stable online identification of the knee point on the predicted degradation curve. The details of the online knee point identification for the predicted degradation curve can be found in Figures S5(c) and S5(d).

**Multi-Task Learning with Sequence-to-Sequence Models**

The input of the MTL model is first pushed to a masking layer to trim the leading zero paddings, ensuring the model learns from only the actual data. A concatenate layer is used to merge the two input sequences into one sequence. In the MTL model, the encoder and hidden vector are shared by the capacity and power degradation prediction tasks. Two different decoders with fully connected networks are used to generate the prediction trajectories for capacity and power degradation. Both encoder and decoder are based on stacked bidirectional long short-term memory (LSTM) networks, which increase the depth of the model and therefore increase the generalization ability and model performance [45]. The equations that govern the operation of the MTL model based on S2S learning are given below,

$$c = LSTM_e\big(Con\big(x_{1,t}, x_{2,t}\big), h_{t-1}\big) \tag{1}$$

$$o_{1,t} = LSTM_{1,d}(c, h_{t-1}) \tag{2}$$

$$o_{2,t} = LSTM_{2,d}(c, h_{t-1}) \tag{3}$$

where $c$ is a hidden vector, $x_{1,t}$ and $x_{2,t}$ are two input sequences, $h_{t-1}$ is the previous hidden state, $o_{1,t}$ and $o_{2,t}$ are the output sequences, $Con(\cdot)$ is the concatenate function, $LSTM_e$, $LSTM_{1,d}$ and $LSTM_{2,d}$ represent the bidirectional LSTM networks in the shared encoder and two decoders, respectively. Although a complex network structure is able to solve complicated problems, the risk of over-fitting for such models in the training process is higher than that of simple network structures. In this work, two regularization methods, L1 and L2, were used to build a sparse MTL model as follows,

$$W_1 = \lambda_1 \sum_{i=1}^{N} |w_i| \tag{4}$$

$$W_2 = \lambda_2 \sum_{i=1}^{N} |w_i^2| \tag{5}$$

where $w_i$ is the weight of the network nodes, $\lambda_1 = 1 \times 10^{-5}$ and $\lambda_2 = 1 \times 10^{-4}$ are the regularization index for the L1 and L2 regularization, respectively, which are determined with the trial-and-error method. The L1 regularization will set parts of the minimal weights as zero [46], and the L2 regularization will constrain the maximum value of the weights [47]. By combining the L1 and L2 regularization methods, the robustness of the MTL model is increased.

**Single-Task Learning with Sequence-to-Sequence Models**

The straightforward approach in battery lifetime prediction is to use the capacity sequence or resistance sequence as the input to output the corresponding future sequence. The STL models proposed in this work are both based on the S2S learning structure, which includes three main parts: an encoder, a decoder, and fully connected layers, as shown in Figure S6. The STL model for capacity degradation prediction shares the same structure as the model we proposed in our previous work in Ref. [26]. The four-layer stacked LSTM bidirectional network was chosen as the encoder and decoder structure to capture the series features. The hidden vector represents the last layer output of the encoder and a repeat-vector layer was used to convert the two-dimensional vector from the encoder to a three-dimensional vector as the input to the decoder. The output length of the decoder was designed up to output all possible future degradation arrays in one shot through the fully connected network with three layers.

**Model Training**

In this work, the MTL-S2S architecture was developed in a Python 3.7 environment, with TensorFlow [48] as the backend and the Keras deep learning library [49] for layer creation. The whole dataset was divided into training data, validation data, and test data with the ratio of 6:2:2, randomly. The training data and validation data are used to optimize the model parameters, and the test data is used to evaluate the model performance. To reduce the computational cost of the model, we resample the input and output data so that every time step represents five cycles and 20 cycles, respectively. Both the encoder and two decoders consist of four bidirectional LSTM layers, and all LSTM layers consist of 64 nodes. The number of nodes in the input layer and the output layers were determined to cover all possible cycle number values in the input and output within the ageing dataset. The final node number of the input and two output layers are 384, 128, and 128, respectively. As the input and output sequences have different lengths at different lifetimes of the cells, the training data was padded before being forwarded to the MTL model (see Table S1 for training parameters of the MTL model).

Training of the STL models is usually based on one loss function considering the output prediction error to update the model parameters. In contrast, the training process of the proposed MTL model is a non-convex optimization problem [50], which is a more challenging task, as there are two loss functions regarding the capacity and power prediction errors. Different losses may update the weights in the shared encoder into different directions, which leads to the slow convergence of the model. To solve this problem and accelerate the training, a multi-stage training process is developed. First, the capacity prediction loss was used to update the weights in the shared encoder and the capacity prediction decoder. Second, the weights in the resistance prediction decoder were updated based on the resistance prediction loss. In the final training stage, the weights of the decoders were updated based on their own loss, but the weights in the shared encoder were updated based on the total loss of the model, which is a weighted sum of the two losses. A modified mean absolute error (MAE) which can skip zero-padding positions, was used as the loss function for both capacity and power degradation prediction. The optimizer used in this work, "Adam", is the state-of-the-art optimizer to train the time-series NNs. An early stopping method was used to avoid overfitting. When the training process shows no improvement of more than a threshold, this method can terminate the process.

**Evaluation Criteria**

The evaluation metrics for both capacity and power degradation prediction are based on the mean absolute percentage errors (MAPEs) as follows,

$$APE = \left| \frac{\hat{y} - y}{y} \right| \times 100\% \tag{6}$$

$$MAPE = \frac{1}{n} \sum_{i=1}^{n} \text{APE}_i \tag{7}$$

which represent the mean deviation based on each prediction point in the trajectory, $y$ and $\hat{y}$ are the real value and predicted value of the output, respectively, and $n$ is the number of the prediction points. The mean, median, and maximum of the MAPE of all the curves from the entire lifetime of the cells are taken as representative metrics. Apart from the curve MAPEs, the absolute cycle errors of EOLs for the first- and second-life and knee-point are also looked at and compared for both capacity and power degradation prediction. These additional metrics could help to evaluate the performance of the model lifetime prediction comprehensively.

**ACKNOWLEDGEMENTS**

This work has received funding from the research project "Model2life" (03XP0334) funded by the German Federal Ministry of Education and Research (BMBF). We would like to thank S. Bihn from RWTH Aachen University and W. Chueh from Stanford University for their review and discussions. We also appreciate the work of M. Zheng from RWTH Aachen University in data visualization.

**AUTHOR CONTRIBUTIONS**

Conceptualization, W.L., H.Z. and D.U.S.; Methodology, W.L., H.Z. and B.V.V.; Invstigation, W.L., H.Z., B.V.V., P.D. and D.U.S.; Writing – Original Draft, W.L. and H.Z.; Writing – Review & Editing, W.L., B.V.V. and D.U.S.; Funding Acquisition, D.U.S.; Supervision, W.L. and D.U.S.

**DECLARATION OF INTERESTS**

W.L. and D.U.S. have filed a patent related to this work: DE Application No. 102020210148.5.

# Forecasting battery capacity and power degradation with multi-task learning

Weihan Li[1,2*] , Haotian Zhang[1], Bruis van Vlijmen[4,5], Philipp Dechent[1,2], Dirk Uwe Sauer[1,2,3*]

1 Chair for Electrochemical Energy Conversion and Storage Systems, Institute for Power Electronics and Electrical Drives (ISEA), RWTH Aachen University, Jägerstraße 17-19, 52066 Aachen, Germany
2 Jülich Aachen Research Alliance, JARA-Energy, Templergraben 55, 52056 Aachen, Germany
3 Helmholtz Institute Münster (HI MS), IEK-12, Forschungszentrum Jülich, 52425 Jülich, Germany
4 SLAC National Accelerator Laboratory, Menlo Park, California 94025, United States of America
5 Stanford University, Stanford, California 94305, United States of America

*Correspondence: weihan.li@isea.rwth-aachen.de (W.L.), dirkuwe.sauer@isea.rwth-aachen.de (D.U.S.)

**Supplemental Items:**

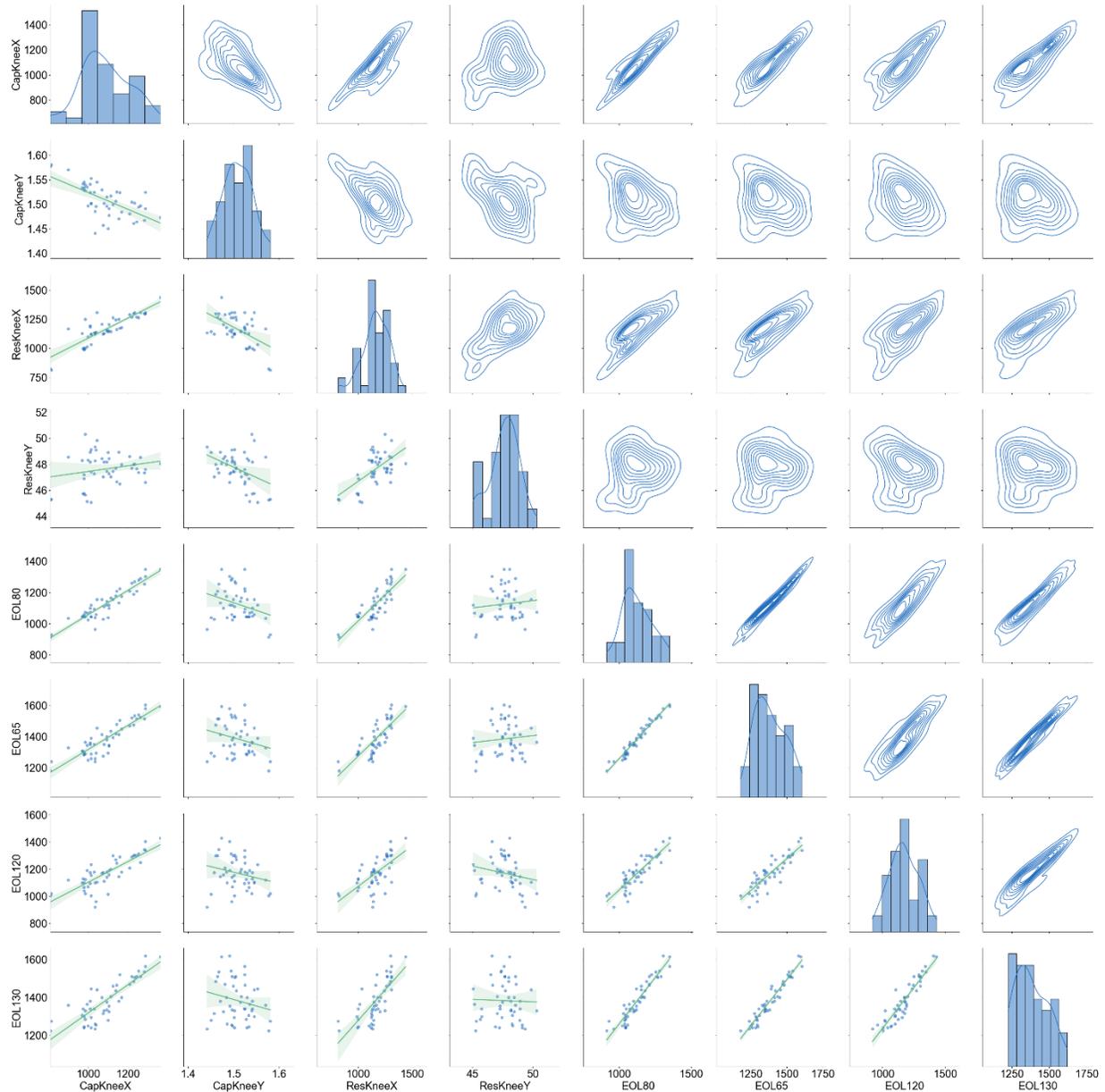

**Figure S1.** Pairwise correlation plot for the metrics of capacity degradation and power degradation. From left to right: 1) CapKneeX: Capacity knee-point in cycles, 2) CapKneeY: Capacity knee-point in Ah, 3) ResKneeX: Resistance knee-point in cycles, 4) ResKneeY: Resistance knee-point in $m\Omega$, 5) EOL80: EOL80 in cycles, 6) EOL65: EOL65 in cycles, 7) EOL120: EOL120 in cycles, 8) EOL130: EOL130 in cycles.

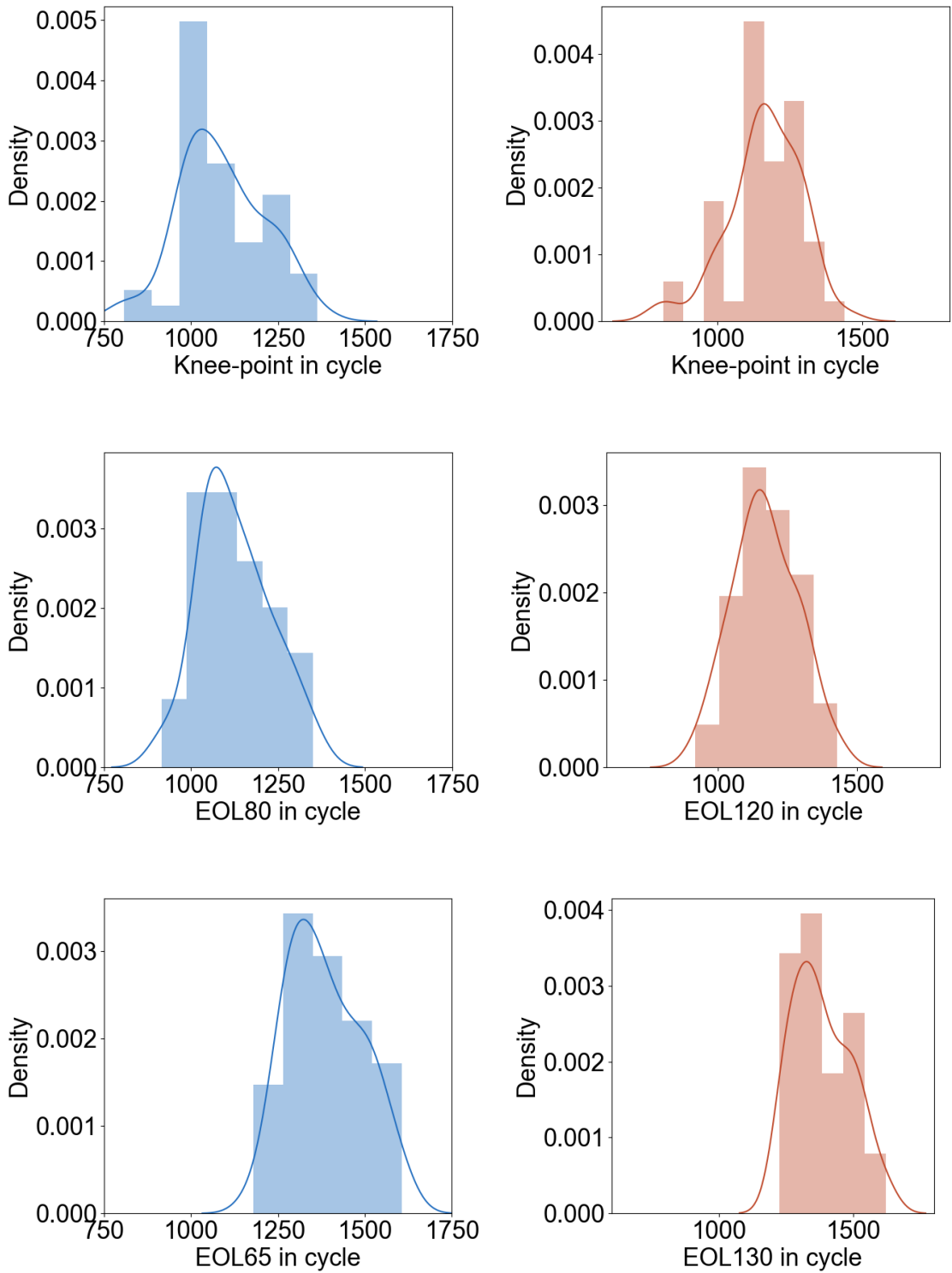

**Figure S2.** The distribution of the capacity degradation metrics (a) capacity knee-point (c) EOL80 and (e) EOL65 and power degradation metrics (b) resistance knee-point (d) EOL120 and (f) EOL130, as determined from the experimental data of 48 NMC/Graphite cells.

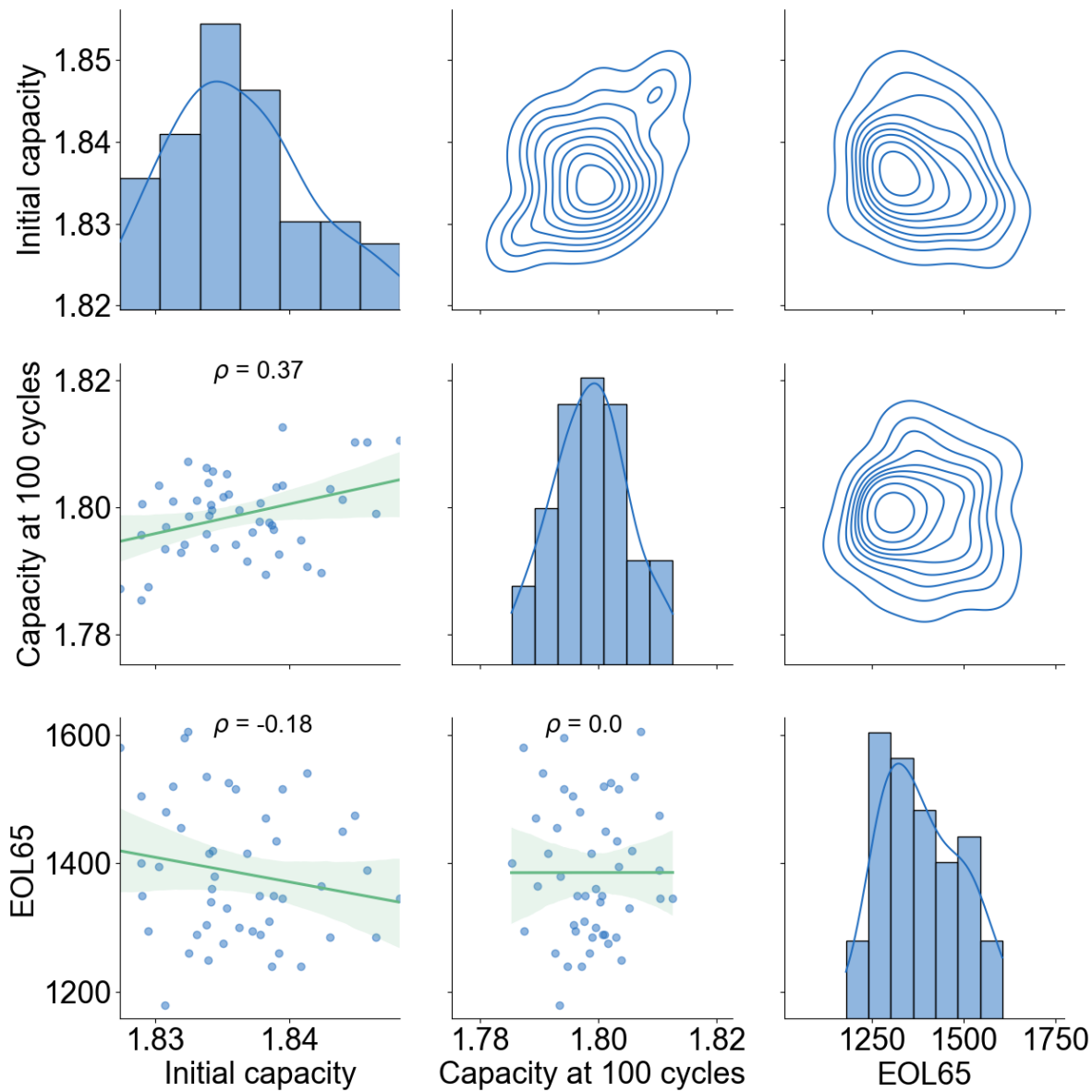

**Figure S3.** Pairwise correlation plot for the initial capacity, remaining capacity at 100 cycles and the battery life cycle at EOL80 of 48 cells. As shown by the Pearson correlation coefficient, both the initial capacity and the capacity at 100 cycles have a very weak correlation with the EOL65.

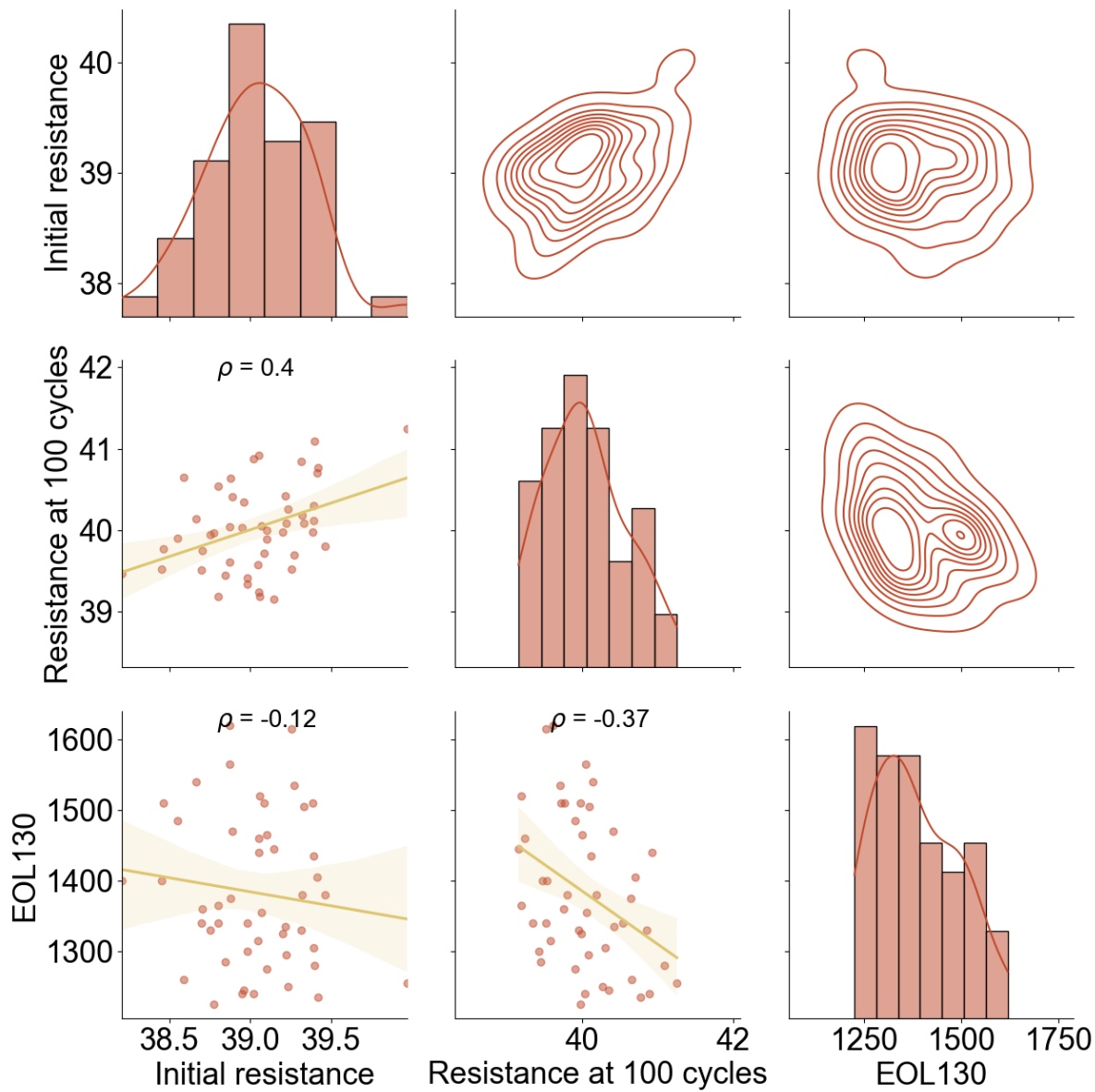

**Figure S4.** Pairwise correlation plot for the initial resistance, resistance at 100 cycles and the battery life cycle at EOL130 of 48 cells. Similar to the capacity degradation, the initial resistance and resistance at 100 cycles show a weak correlation with EOL130. However, the correlation between the resistance at 100 cycles with EOL130 ($\rho = -0.37$) is much stronger than that between the capacity at 100 cycles with EOL65 ($\rho = 0$), as shown in Figure S3.

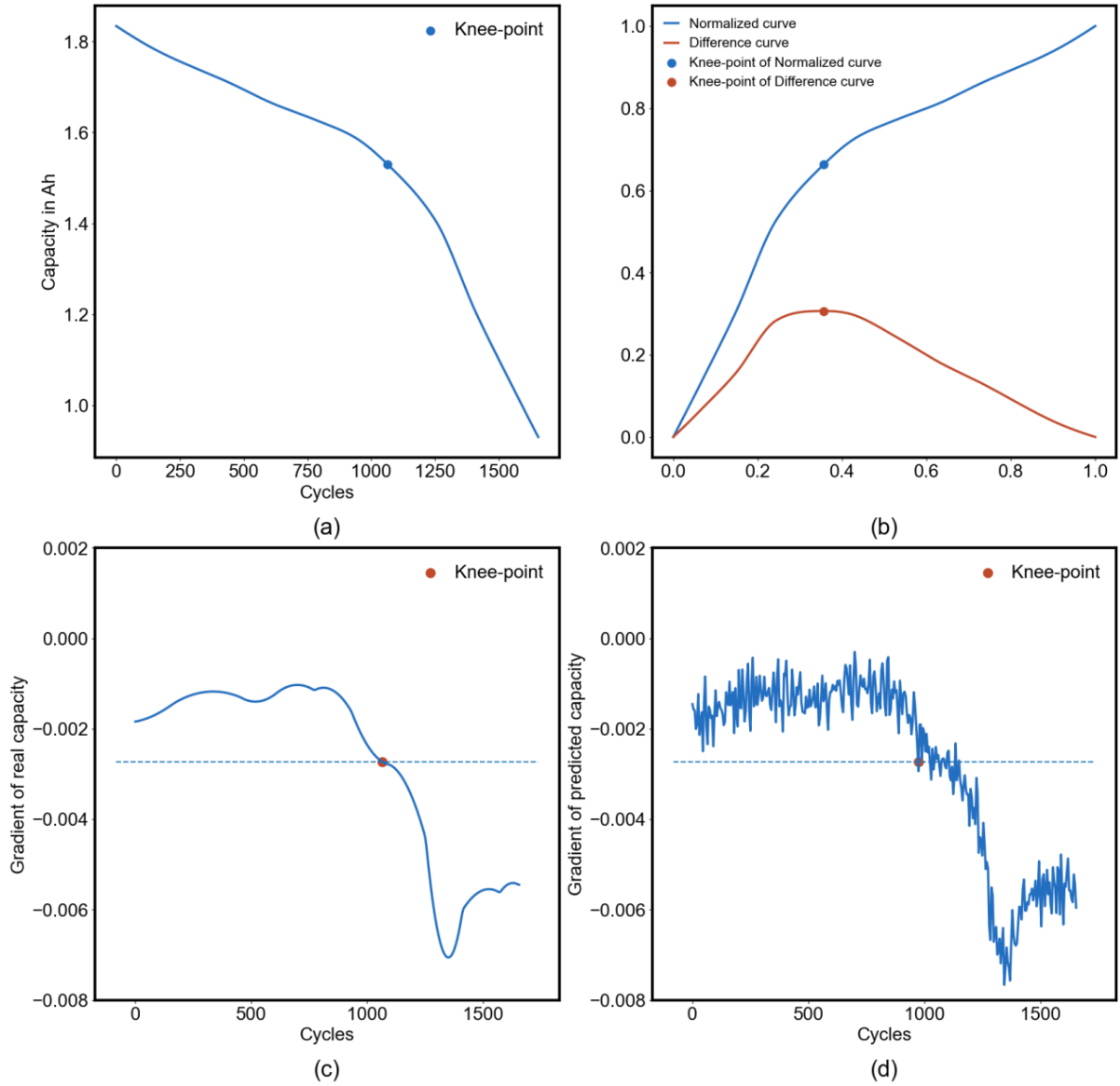

**Figure S5.** Knee-point identification method. (a) Original degradation data curve. (b) Normalized degradation data curve and x-y-difference curve. (c) Gradient curve of the reference curve. (d) Gradient curve of the predicted curve.

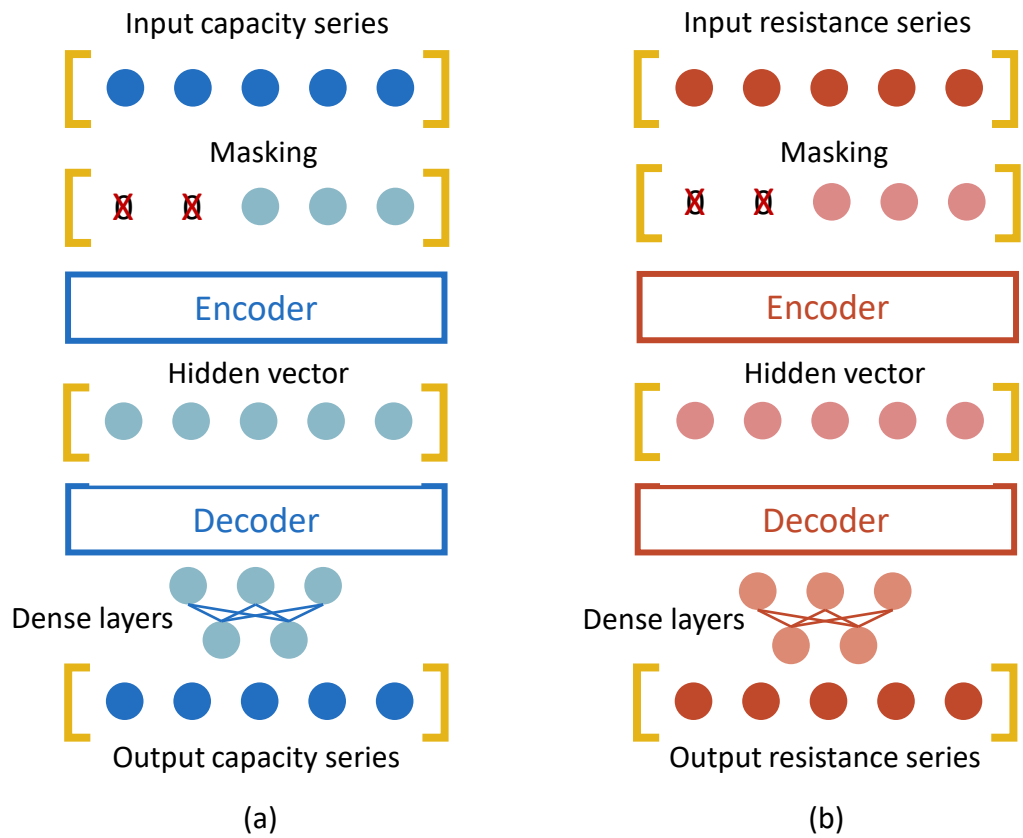

**Figure S6.** (a) The architecture of the single-task sequence-to-sequence learning model for capacity degradation prediction. (b) The architecture of the single-task sequence-to-sequence learning model for power degradation prediction.

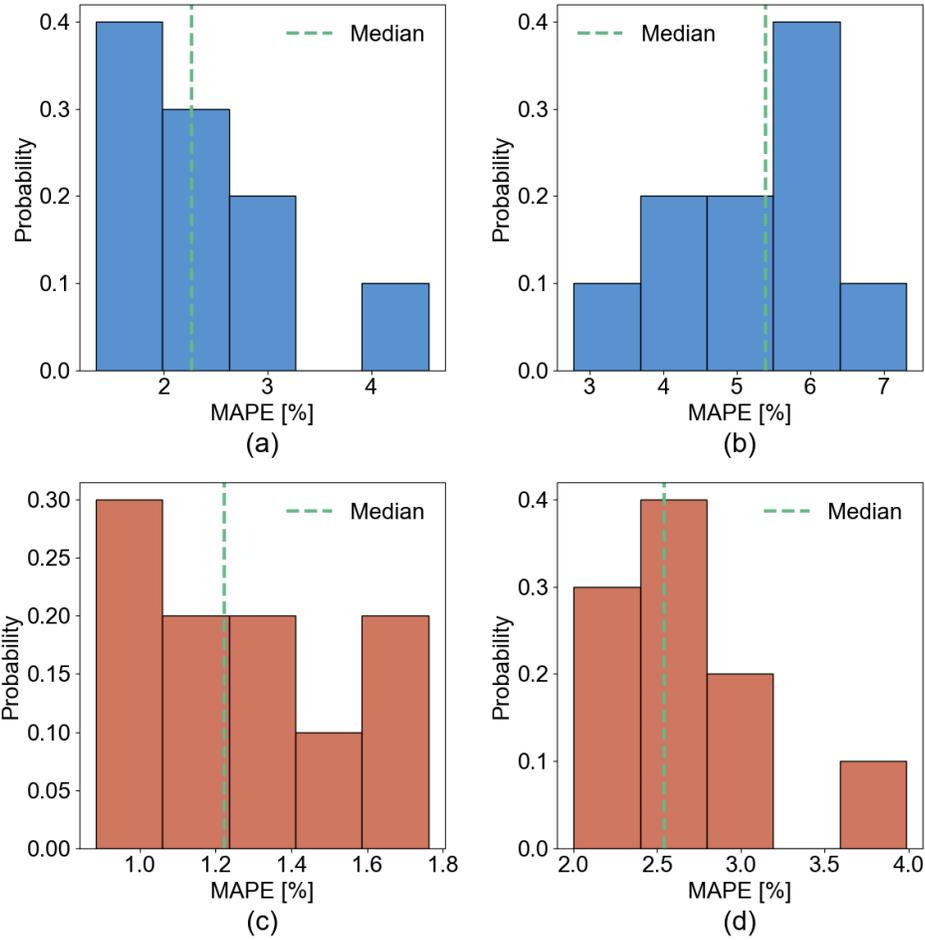

**Figure S7.** MAPE distribution of the MTL model under the normal condition: (a) Mean capacity prediction MAPE distribution. (b) Maximum capacity prediction MAPE. (c) Mean internal resistance prediction MAPE distribution. (d) Maximum internal resistance prediction MAPE.

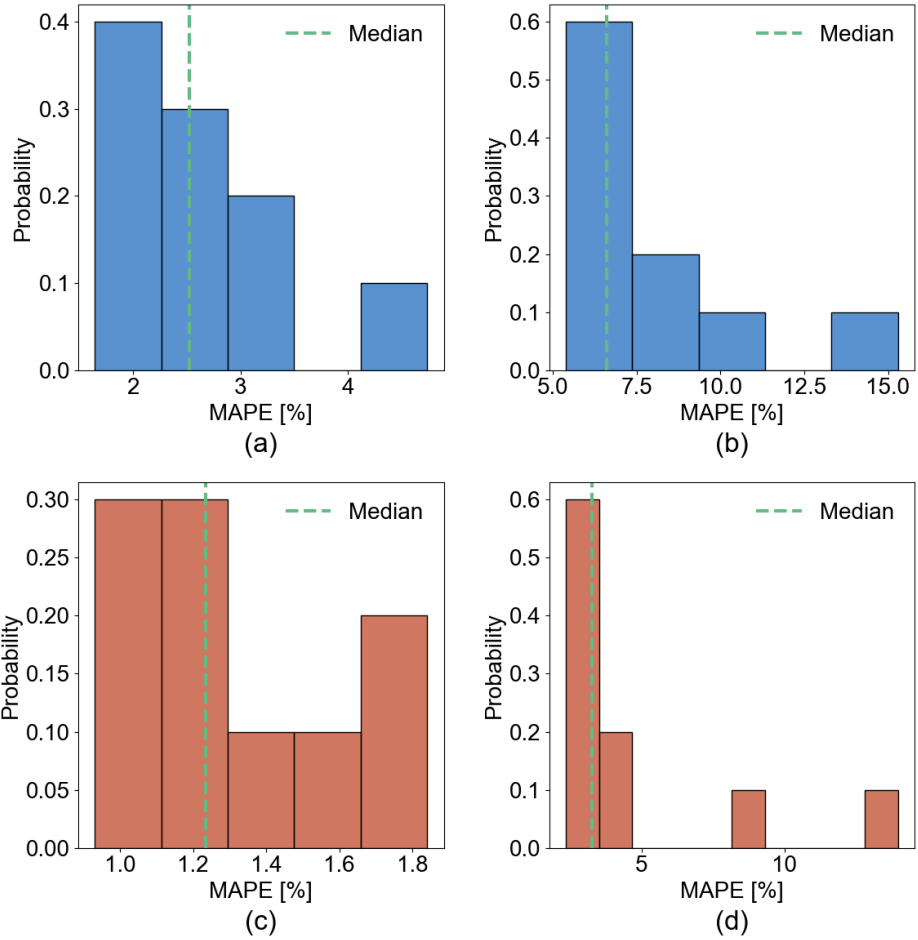

**Figure S8.** MAPE distribution of the MTL model under the noisy input condition: (a) Mean capacity prediction MAPE distribution. (b) Maximum capacity prediction MAPE. (c) Mean internal resistance prediction MAPE distribution. (d) Maximum internal resistance prediction MAPE.

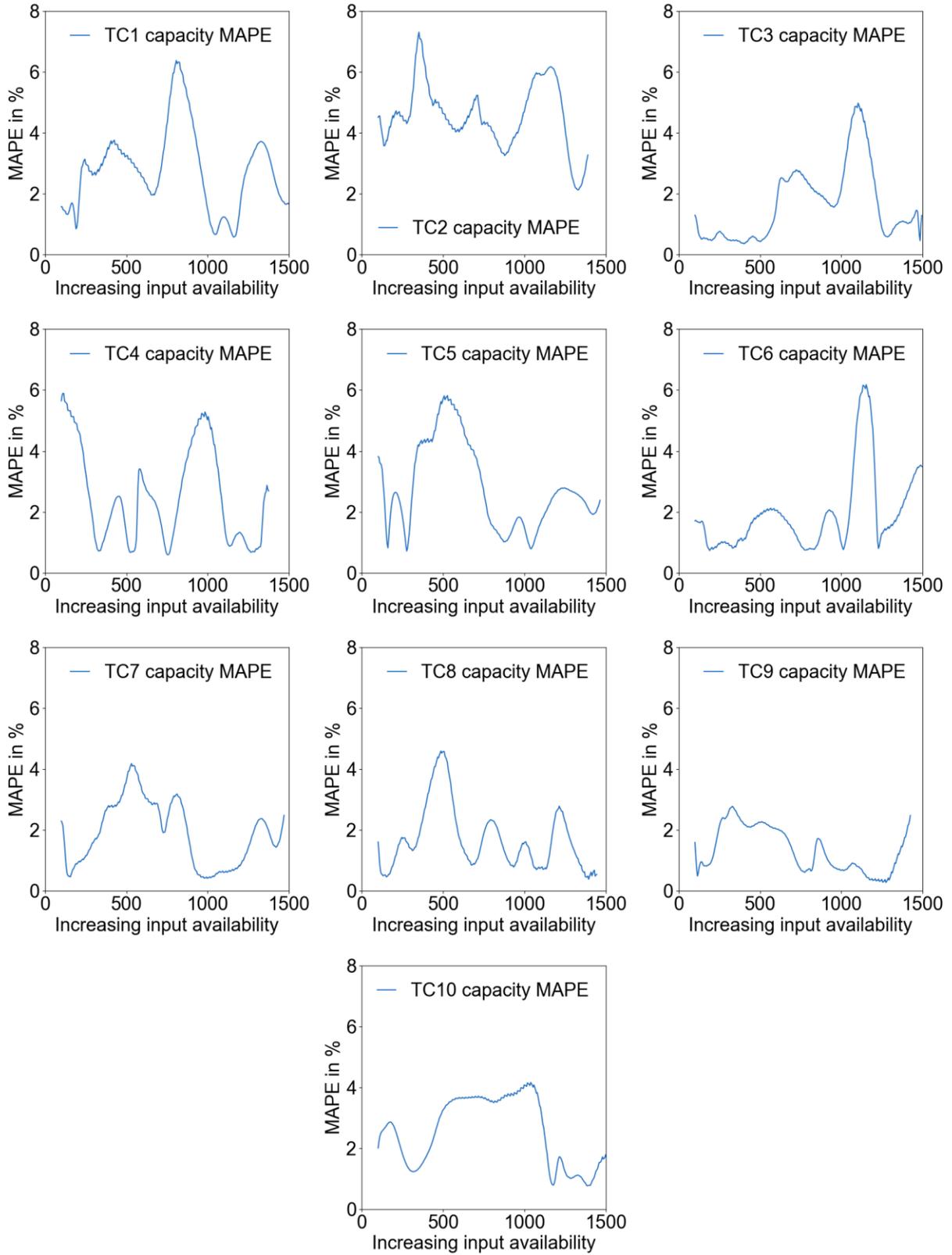

**Figure S9.** Capacity fade trajectory prediction MAPE of the MTL model for each cell in the test set (10 cells).

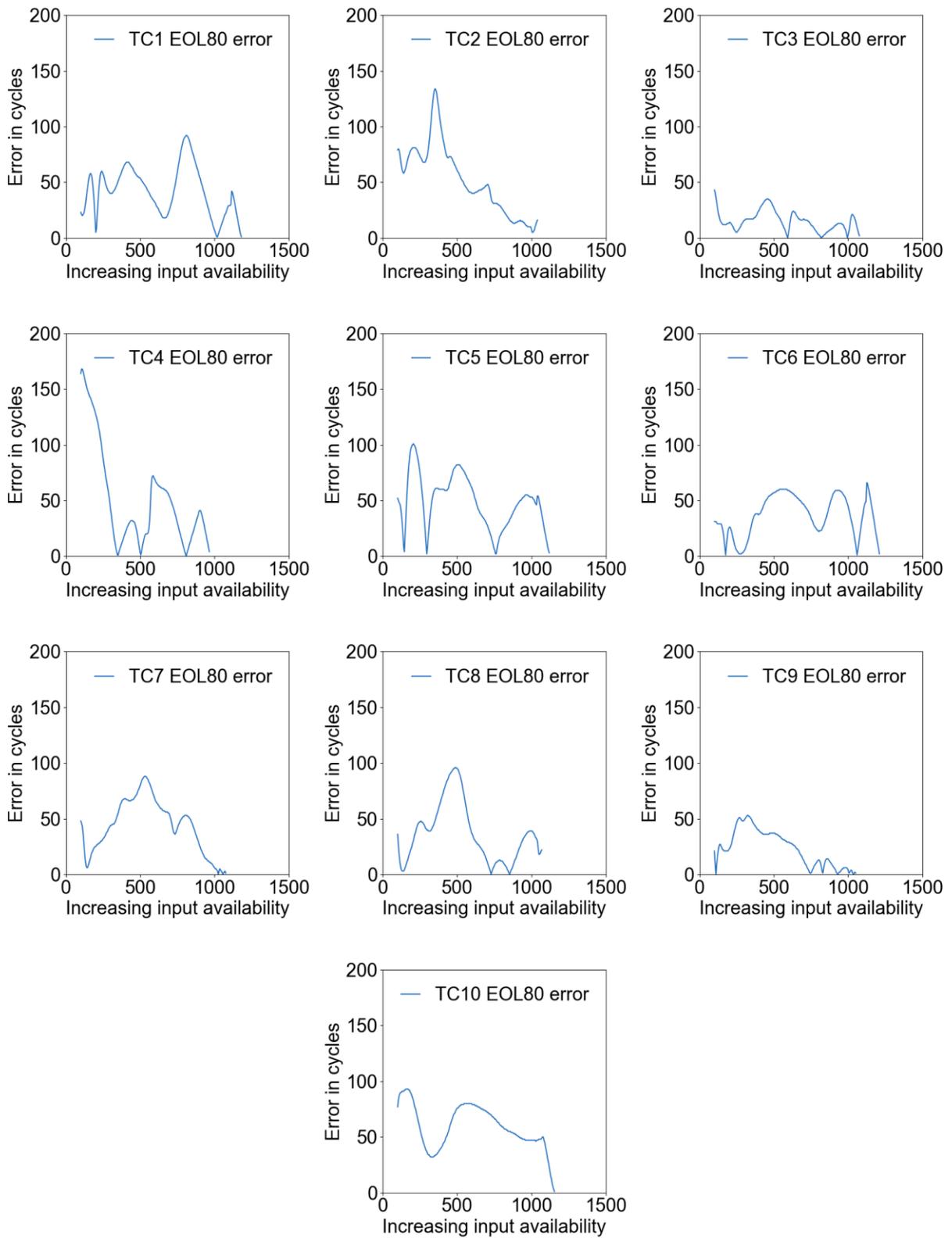

**Figure S10.** Capacity EOL80 prediction error of the MTL model for each cell in the test set (10 cells).

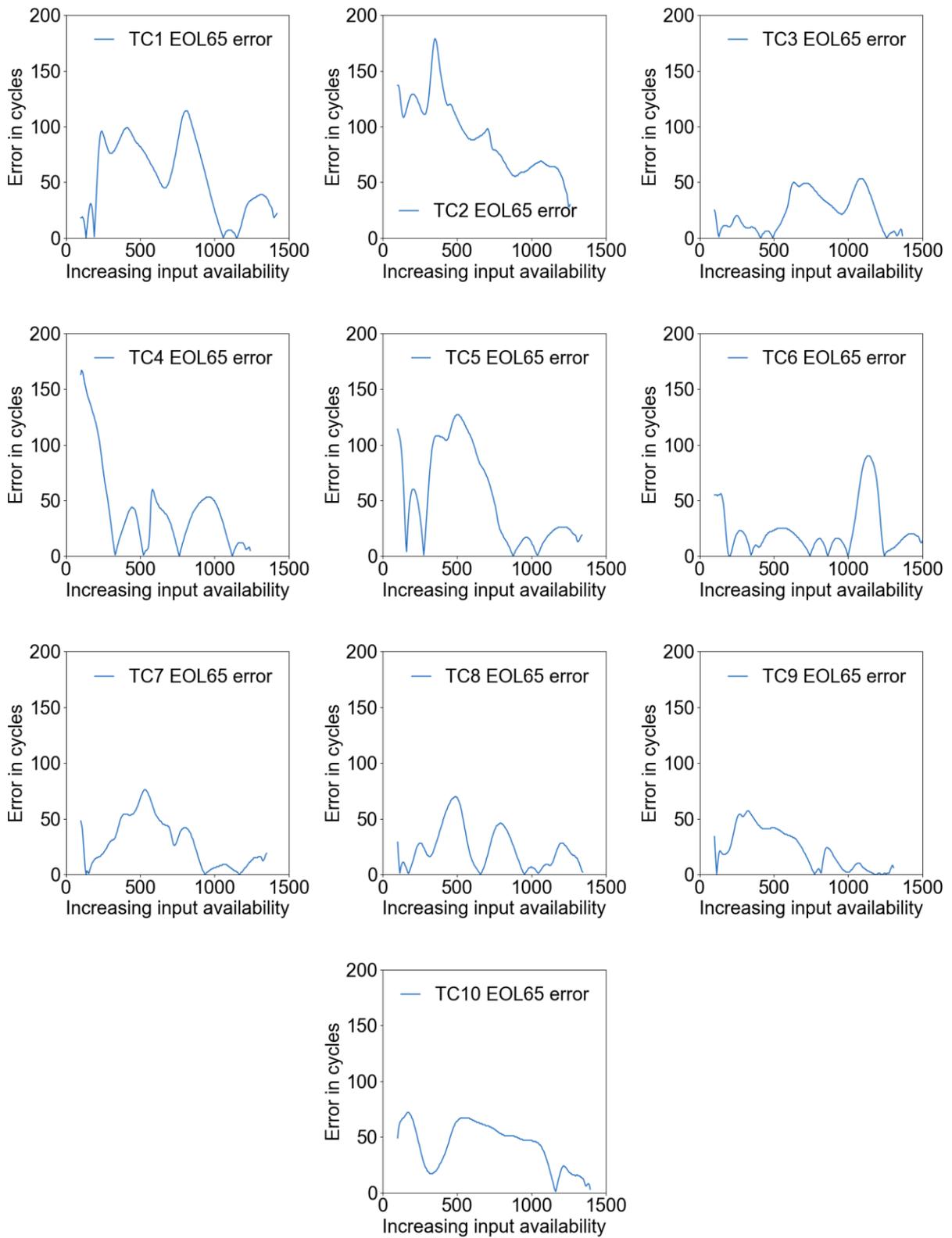

**Figure S11.** Capacity EOL65 prediction error of the MTL model for each cell in the test set (10 cells).

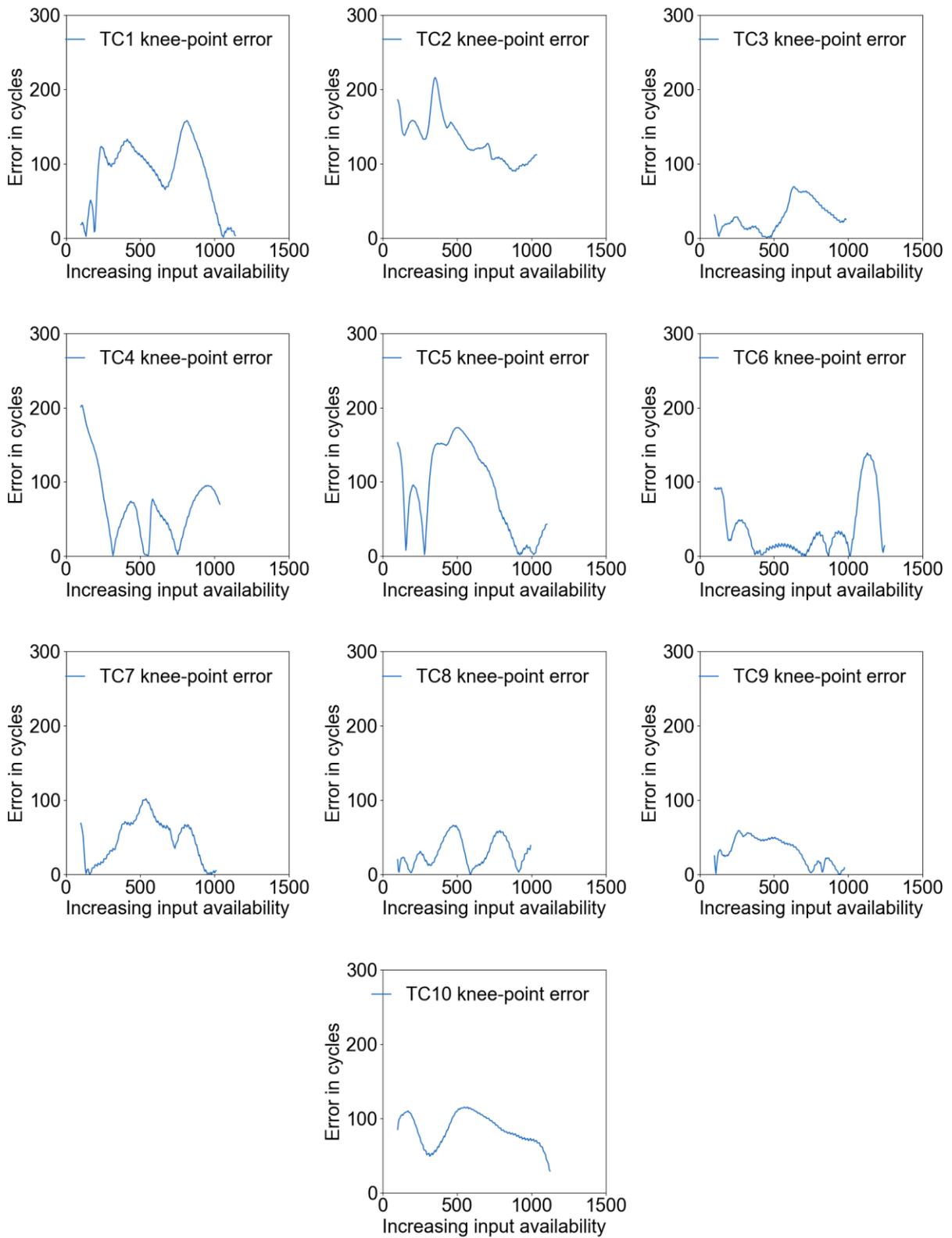

**Figure S12.** Capacity knee-point prediction error of the MTL model for each cell in the test set (10 cells).

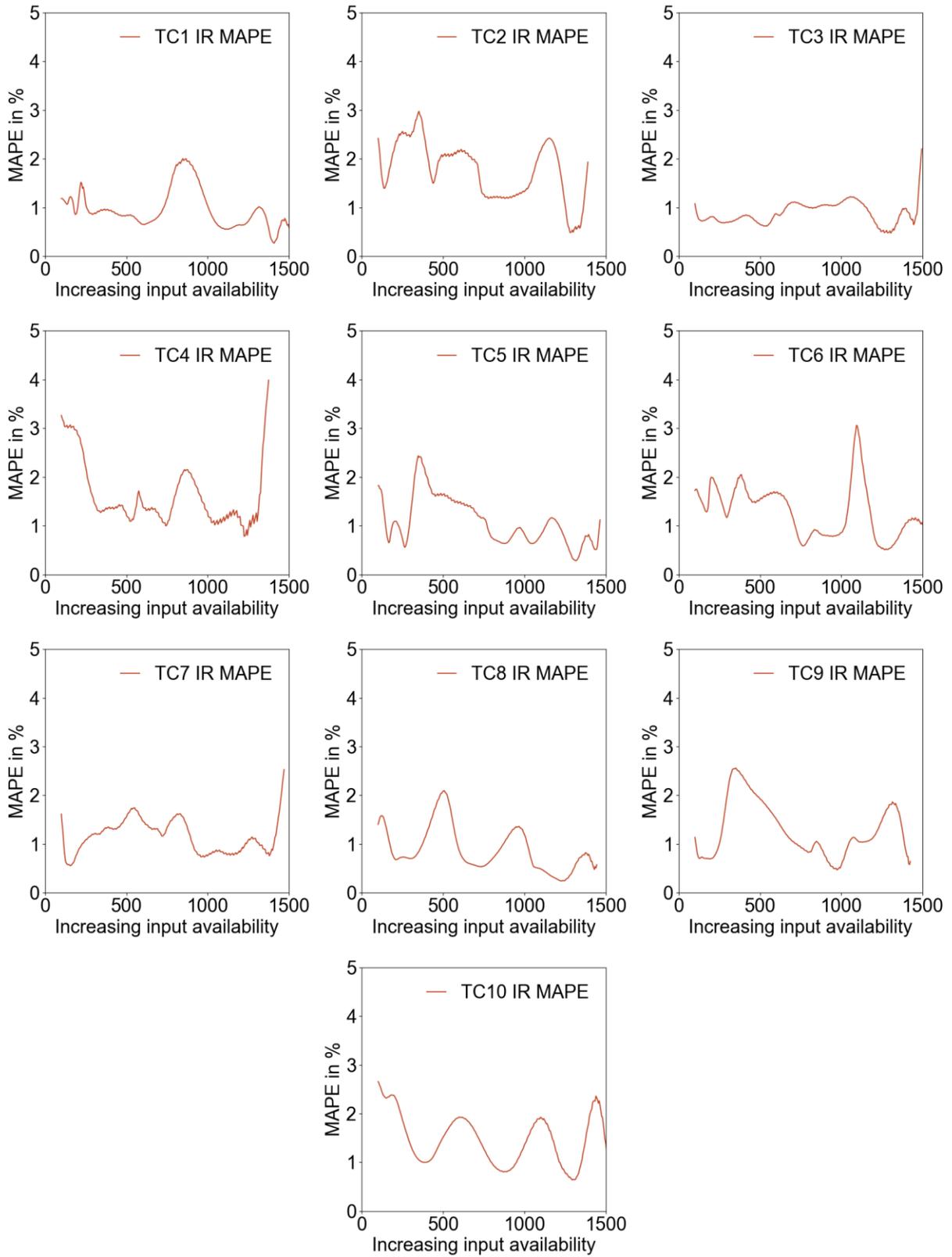

**Figure S13.** Power degradation trajectory prediction MAPE of the MTL model for each cell in the test set (10 cells).

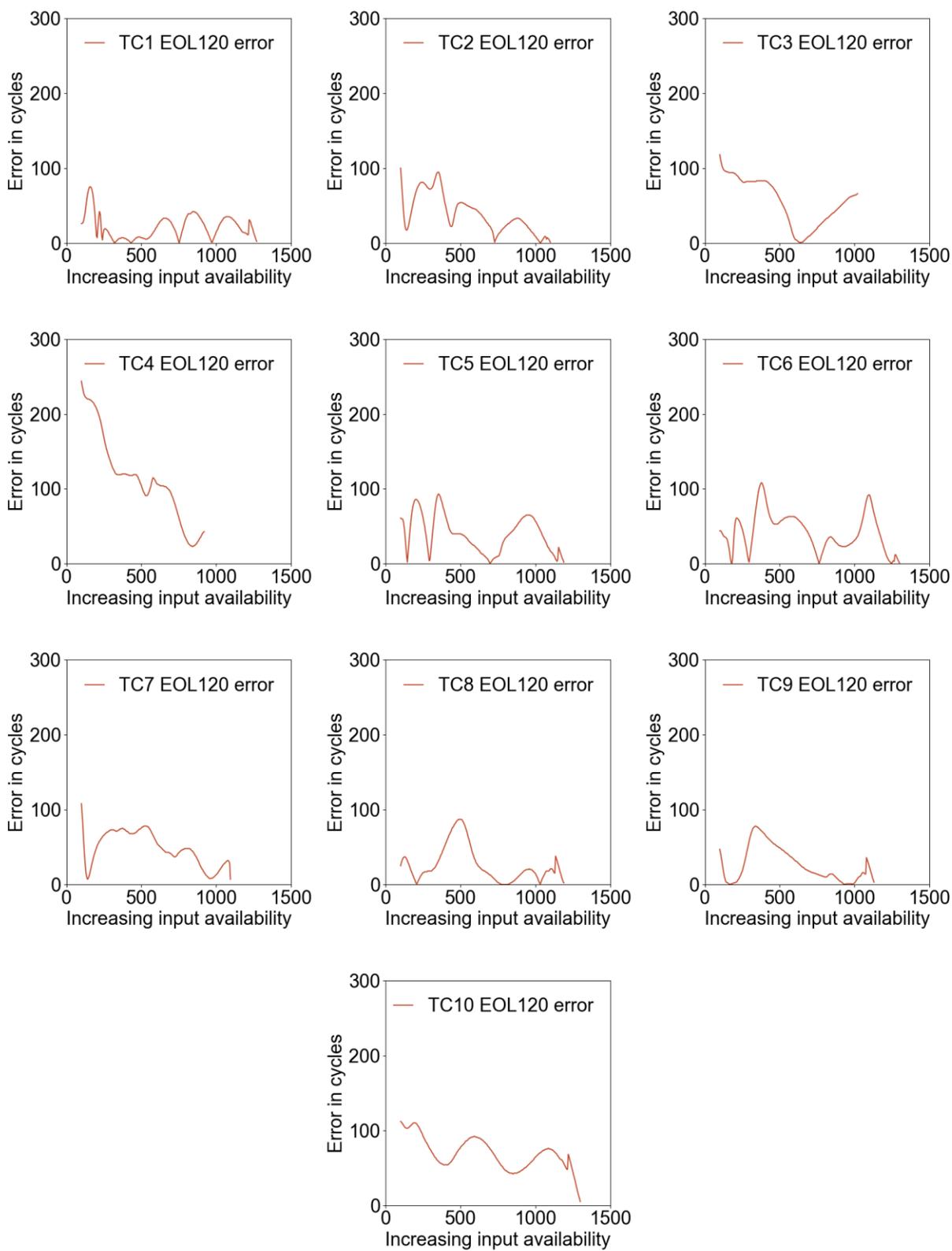

**Figure S14.** Resistance EOL120 prediction error of the MTL model for each cell in the test set (10 cells).

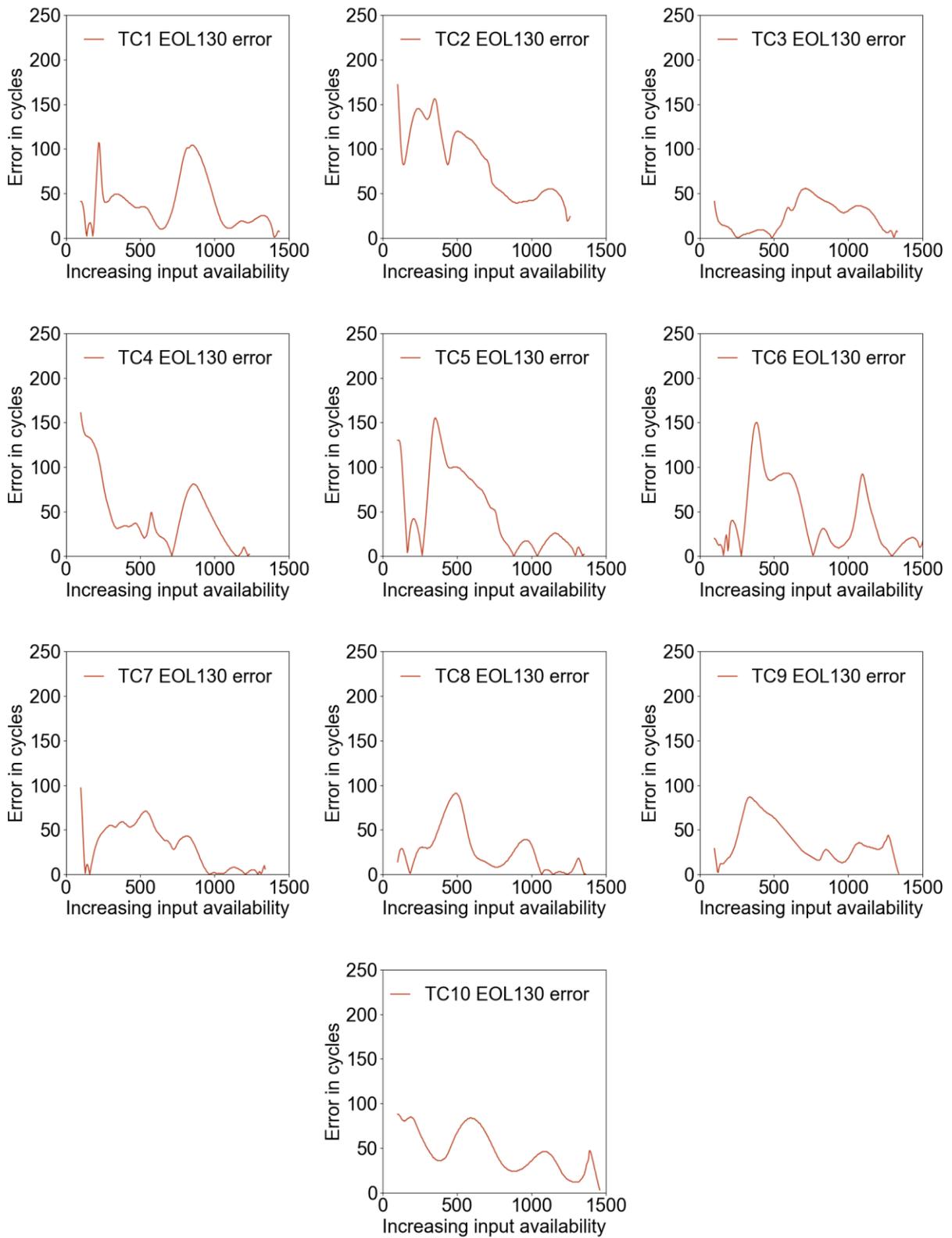

**Figure S15.** Resistance EOL130 prediction error of the MTL model for each cell in the test set (10 cells).

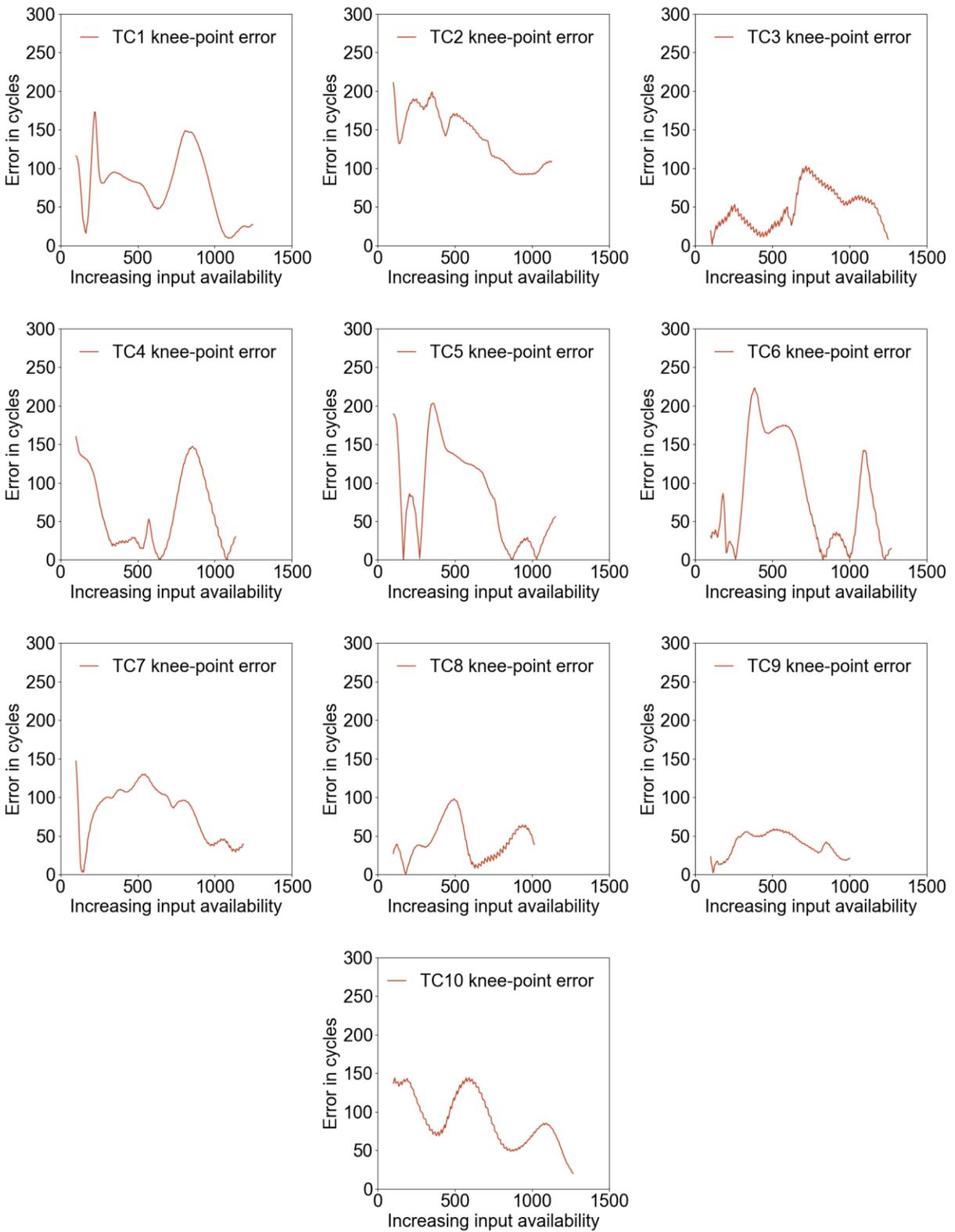

**Figure S16.** Resistance knee-point prediction error of the MTL model for each cell in the test set (10 cells).

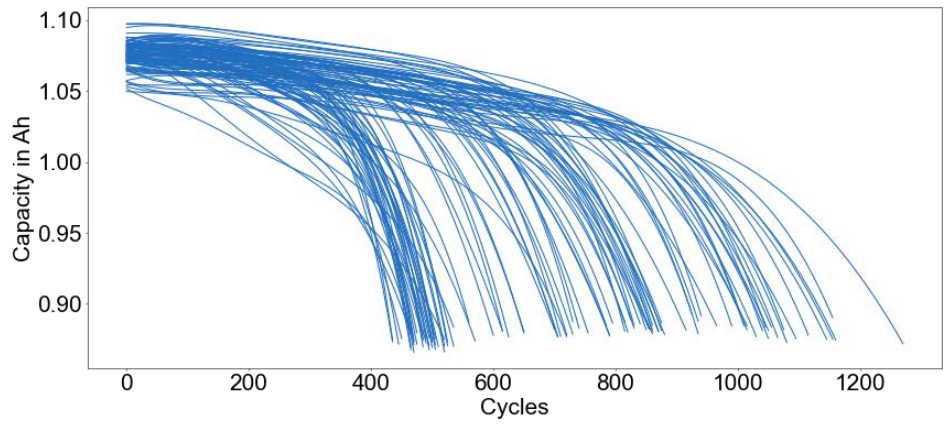

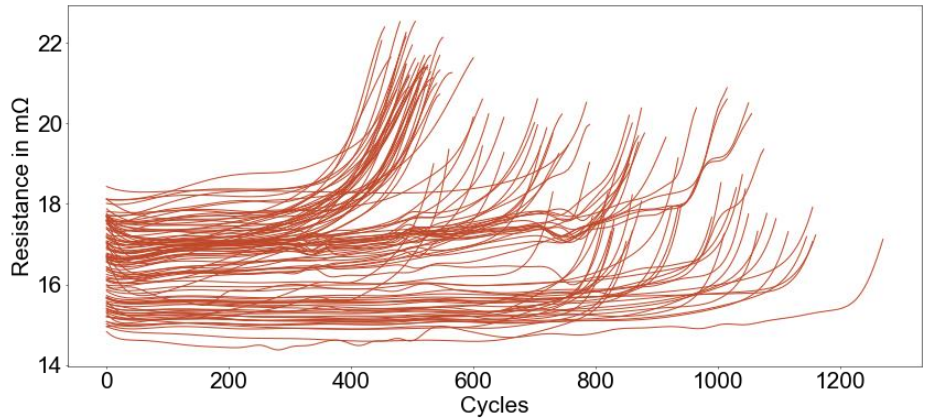

**Figure S17.** The battery degradation dataset by Severson et al. shows large degradation variations due to the differences in external stress factors. (a) Capacity fade data. (b) Resistance increase data.

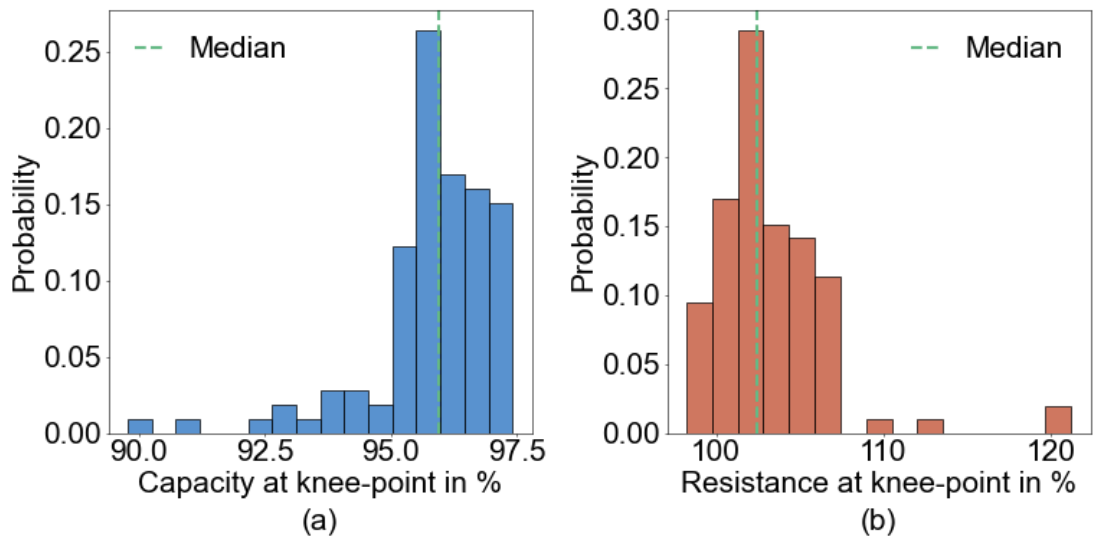

**Figure S18.** The battery degradation dataset by Severson et al. shows large degradation variations due to the differences in external stress factors. (a): Distribution of the remaining capacity where the capacity knee points are reached by the cells. (b) Distribution of the internal resistance where the resistance knee points are reached by the cells.

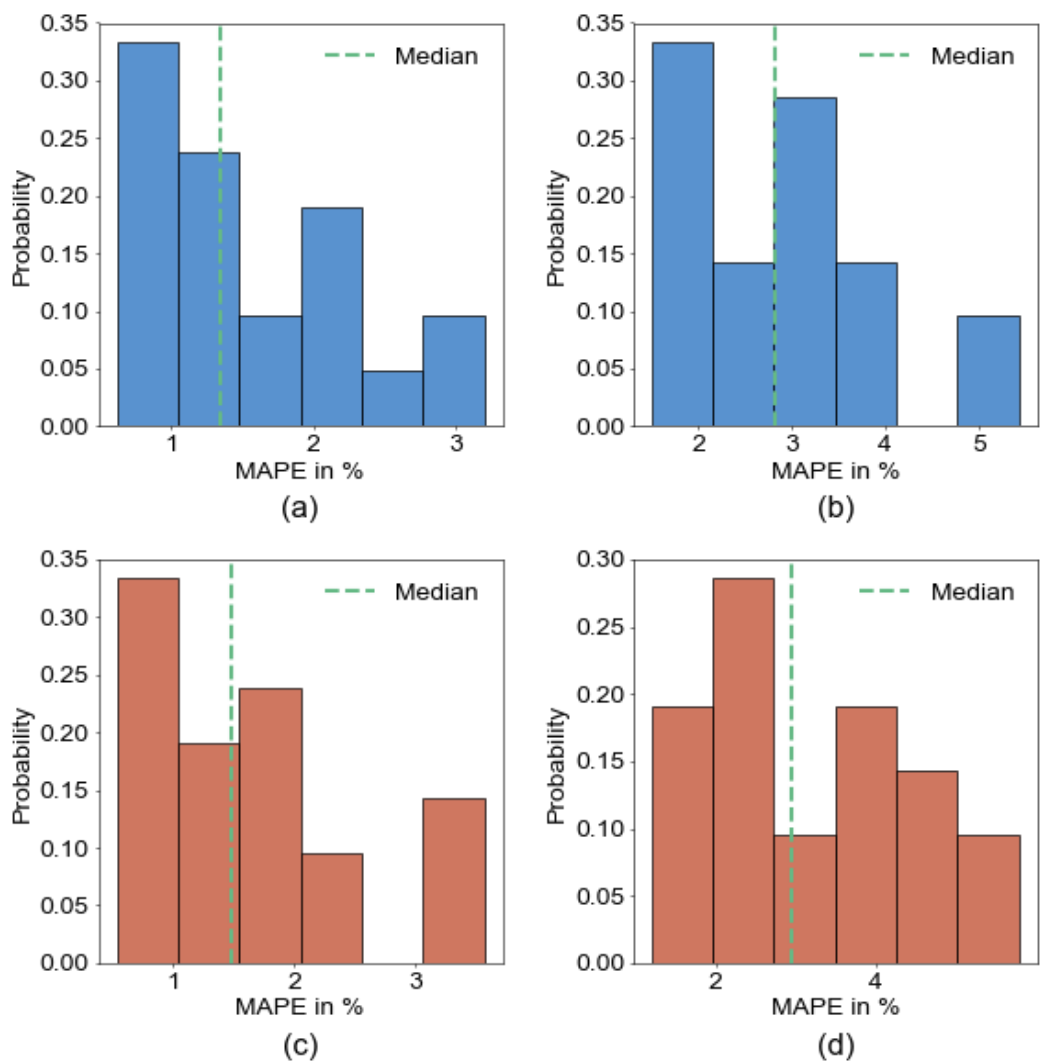

**Figure S19.** MTL model performance with the Severson dataset: (a) Distribution of the mean capacity fade prediction MAPE. (b) Distribution of the maximum capacity fade prediction MAPE. (c) Distribution of the mean internal resistance prediction MAPE. (b) Distribution of the maximum internal resistance prediction MAPE.

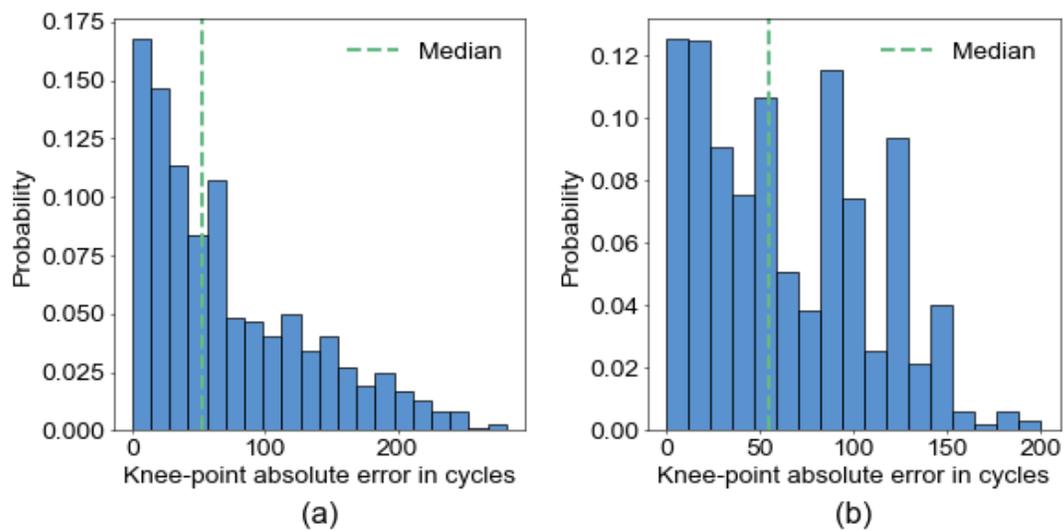

**Figure S20.** MTL model performance with the Severson dataset: (a) Distribution histogram of the absolute errors in knee-point prediction for the battery capacity fade. (b) Distribution histogram of the absolute errors in knee-point prediction for the battery resistance increase.

**Table S1.** MTL model training parameters.

| Category | Stage 1 | Stage 2 | Stage 3 |
|---|---|---|---|
| Optimizer | Adam | Adam | Adam |
| Learning rate | 1e-4 | 1e-4 | 1e-5 |
| Early stopping | 32 epochs | 32 epochs | 32 epochs |
| Epoch | 450 | 450 | 300 |
| Batch size | 384 | 384 | 512 |
| Loss weight | [1.0,0.0] | [0.0,1.0] | [1.0,1.0] |
| Loss function | Capacity MAE | Resistance MAE | Total MAE |

**Table S2.** STL model training parameters.

| Category | Capacity degradation model | Power degradation model |
|---|---|---|
| Optimizer | Adam | Adam |
| Learning rate | 1e-4 | 1e-4 |
| Early stopping | 32 epochs | 32 epochs |
| Epoch | 450 | 450 |
| Batch size | 384 | 384 |
| Loss function | Capacity MAE | Resistance MAE |

**Table S3.** Comparison between the MTL model and STL models.

|  | STL | MTL |
|---|---|---|
| Mean capacity curve MAPE [%] | 2.85 | 2.37 |
| Median capacity curve MAPE [%] | 2.83 | 2.27 |
| Max capacity curve MAPE [%] | 8.25 | 7.30 |
| Median capacity knee-point error [Cycle] | 73 | 55 |
| Median EOL80 error [Cycle] | 39 | 38 |
| Medan EOL65 error [Cycle] | 34 | 30 |
| Mean resistance curve MAPE [%] | 1.45 | 1.24 |
| Median resistance curve MAPE [%] | 1.38 | 1.22 |
| Max resistance curve MAPE [%] | 4.51 | 3.99 |
| Median resistance knee-point error [%] | 84 | 64 |
| Median EOL120 error [Cycle] | 39 | 41 |
| Median EOL130 error [Cycle] | 35 | 34 |
| Mean computational cost [s] | 0.34 | 0.17 |